\documentclass[aps,twocolumn,prd,superscriptaddress,nofootinbib,floatfix]{revtex4-2}

\usepackage{mathtools}
\usepackage{amsfonts}
\usepackage{amssymb}
\usepackage{mathrsfs}
\usepackage{bbm}
\usepackage{slashed}
\usepackage{amsmath}
\usepackage{bm}
\usepackage{bbold}

\usepackage{graphicx}
\usepackage{color}
\usepackage{colortbl}
\usepackage{array}

\usepackage{float}
\usepackage{placeins}
\usepackage{booktabs}
\usepackage[caption=false]{subfig}
\usepackage{makecell}
\usepackage{tabstackengine}

\usepackage{xspace}
\usepackage{hyperref}
\usepackage[nameinlink]{cleveref}
\usepackage{bookmark}
\usepackage{siunitx}

\usepackage{xifthen}
\usepackage{xcolor}
\hypersetup{
	colorlinks,
	linkcolor={red!75!black},
	citecolor={blue!75!black},
	urlcolor={blue!75!black}
}

\usepackage[utf8]{inputenc}
\allowdisplaybreaks[4]

\newcommand{\feyn}[1]{
	\setbox0=\hbox{\ensuremath{#1}}
	\hbox to\wd0{\hbox to0pt{\hbox to\wd0{\hss/\hss}\hss}\box0}}

\def\Eq#1{Eq.~\labelcref{#1}}

\def\Fig#1{Fig.~\labelcref{#1}}
\def\Tab#1{Tab.~\labelcref{#1}}
\def\sec#1{Sec.~\labelcref{#1}}
\def\app#1{App.~\labelcref{#1}}

\setkeys{Gin}{width=0.48\textwidth}
\graphicspath{{./figures/}}

\newcommand{\gettitle}{}

\newcommand{\getDalianAffiliation}{\affiliation{School of Physics, Dalian University of Technology, Dalian, 116024, P.R. China}}

\newcommand{\getGiessenAffiliation}{\affiliation{Institut f\"ur Theoretische Physik, Justus-Liebig-Universit\"at Gie\ss en, 35392 Gie\ss en, Germany}}

\hypersetup{
	colorlinks,
	linkcolor={red!75!black},
	citecolor={blue!75!black},
	urlcolor={blue!75!black},
	pdftitle={\gettitle},
	pdfauthor={Zelle},
	pdfkeywords={effective theory} {analytic continuation}
	{correlations functions} {hadronization}
	{functional renormalisation group}
	{real time} {bound states}
	bookmarksopen=true,
	bookmarksopenlevel=2,
	bookmarksnumbered=true
}

\begin{document}

\title{Density functional theory of renormalization group in nuclear matter}
	
\author{Yong-rui Chen}
\email{yong-rui.chen@uni-giessen.de}
\getDalianAffiliation
\getGiessenAffiliation

\author{Wei-jie Fu}
\email{wjfu@dlut.edu.cn}
\getDalianAffiliation

\author{Yang-yang Tan}
\email{yangyangtan@mail.dlut.edu.cn}
\getDalianAffiliation

\begin{abstract} 

The density functional renormalization group (density-fRG) is proposed to investigate the density fluctuations within the functional renormalization group approach, which allows us to quantify the medium effect and study physics of high densities. This method is applied to the nucleon-meson effective field theory, also known as the Walecka model, to study the properties of nuclear matter at high baryon densities. It is found that both the attractive and repulsive nucleon meson interactions are screened by the high density medium, which results in a stiffer equation of state (EoS) of nuclear matter in the regime of  $\rho_0 \lesssim \rho \lesssim 2.5 \rho_0$, then a softer EoS when $\rho \gtrsim 2.5 \rho_0$. Here $\rho_0$ denotes the saturation baryon density of symmetric nuclear matter. Furthermore, a new phenomenon called the locking of Fermi surface is found. In the locking of Fermi surface the effective energy of quasi-nucleon is always close to the Fermi surface, which are both running with the renormalization group scale.

\end{abstract}
	
\maketitle

\section{Introduction}
\label{sec:int} 

High density physics of strongly interacting matter has attracted more and more attentions in recent years. On the one hand, it is indicated that the QCD critical end point (CEP) is more probably located in the regime of large baryon chemical potential, i.e., high baryon density, rather than the regime of low baryon chemical potential, in the QCD phase diagram \cite{Stephanov:2007fk, STAR:2010vob, Luo:2017faz, Bzdak:2019pkr, Fu:2022gou, Chen:2024aom} from both recent experiments of relativistic heavy ion collisions \cite{STAR:2020tga, STAR:2021fge, STAR:2022vlo, STAR:2022etb, STAR:2025zdq} and the theoretical studies including first-principles QCD calculations at finite temperature and densities \cite{Fu:2019hdw, Gao:2020fbl, Gunkel:2021oya, Clarke:2024ugt, Sorensen:2024mry}, see also \cite{Bluhm:2024uhj}. On the other hand, the equation of state of strongly interacting matter at high baryon densities is closely related to the astrophysics of compact stars, such as the neutron star and its gravitational wave signatures as well as the multi-messenger astronomy \cite{LIGOScientific:2017vwq, Fujimoto:2019hxv, Koliogiannis:2021ijs, Fujimoto:2022ohj, Fukushima:2025ujk}.

With the rapid progress in new facilities all over the world, there has been a new era of studies of high density physics. The fixed-target experiment within the Beam Energy Scan (BES) program at RHIC has pushed the baryon chemical potential up to $\sim$ 750 MeV \cite{STAR:2021fge, STAR:2022etb}. The Cooling-Storage-Ring External-target Experiment (CEE) at Heavy Ion Research Facility in Lanzhou (HIRFL) in China is designed to study the properties of nuclear matter in the regime of high baryon densities created in heavy-ion collisions \cite{Hu:2023niz}, which will start to operate in 2025. The CBM experiment at FAIR \cite{Friman:2011zz} in Germany will also start to run in the near future. Besides, there are several other heavy-ion facilities planned or underway, such as HADES at GSI in Germany \cite{Agakishiev:2009am}, HIAF in China \cite{Yang:2013yeb}, the NICA/MPD in Russia \cite{Sorin:2011zz}, J-PARC-HI in Japan \cite{Sakaguchi:2017ggo}.

The rapid advance of high density experiments requires reliable theoretical computation in the regime of high baryon chemical potentials. First-principles lattice QCD simulations have made significant progress in the calculations of QCD phase transitions and QCD thermodynamics in the regime of vanishing and small baryon chemical potential \cite{Bellwied:2015rza, Bazavov:2017dus, Bazavov:2017tot, Borsanyi:2018grb, Bazavov:2018mes, Ding:2019prx, Ding:2020xlj, Bazavov:2020bjn, Borsanyi:2020fev, Clarke:2024ugt}, which are, however, usually restricted to the region of $\mu_B/T\lesssim 2 \sim 3$ because of the sign problem in Monte Carlo simulations at finite densities. Here $\mu_B$ and $T$ refer to the baryon chemical potential and temperature. In complement with lattice simulations, functional QCD, e.g., the functional renormalization group (fRG), is also a powerful theoretical method to first-principles QCD at finite temperature and densities \cite{Fu:2019hdw, Braun:2020ada, Braun:2023qak, Fu:2024rto}. Although the truncation and error control of systematics become increasingly difficult, the first-principles QCD calculations within the fRG can be reliably extended to the regime of $\mu_B/T\sim 4$ \cite{Fu:2023lcm}. For more details about the development and application of fRG, see, e.g., \cite{Mitter:2014wpa, Braun:2014ata, Rennecke:2015eba, Cyrol:2016tym, Cyrol:2017ewj, Fu:2016tey, Fu:2021oaw, Fu:2022uow, Fu:2023lcm, Fu:2024ysj, Tan:2024fuq, Ihssen:2024miv, Fu:2025hcm, Zhang:2025ofc, Chen:2025vwl, Gholami:2025afm} for recent progress and \cite{Dupuis:2020fhh, Fu:2022gou} for review.

In the fRG approach, quantum, thermal and density fluctuations are taken into account successively with the evolution of the renormalization group (RG) scale. The quantum fluctuations are also referred to as the fluctuations in vacuum. In order to simplify the calculations at high baryon densities and in the spirit of density functional theory (DFT), in this work we try to separate the density fluctuations from vacuum fluctuations, and only compute the density fluctuations with the RG evolution. Hereafter we call this method the density functional theory of renormalization group (DFT-RG) or the density functional renormalization group (density-fRG). As a first step, in this paper we would like to apply the density-fRG to the nucleon-meson effective field theory which is also known as the Walecka model \cite{Walecka:1974qa}.

This paper is organized as follows: The nucleon-meson effective field theory within the fRG is briefly described in \Cref{sec:EFT}, followed by the discussions about the flow of the effective potential in \Cref{sec:potential} and the Yukawa couplings in \Cref{sec:Yukawa}. Numerical results are presented in \Cref{sec:num}, followed by \Cref{sec:Interpolation}, where we discuss the relation between the fRG and the relativistic mean field calculations. In \Cref{sec:conclusion} we conclude and give an outlook. Some notations are collected in \Cref{app:notat}. The explicit expressions of the flows of Yukawa couplings are presented in \Cref{app:Yukawa}, and the threshold functions used in this work are given in \Cref{app:thres}.

\section{Nucleon meson effective field theory within the fRG}
\label{sec:EFT} 

We begin with the effective action of the nucleon-meson effective field theory (EFT) within the functional renormalization group approach, which reads
\begin{align}
    \Gamma_{k}[\Phi]=&\int_x \bigg\{ \bar{\psi}_N \Big[\gamma_\mu \partial_\mu+m_N-\gamma_0\hat\mu_N-h_{\sigma,k}\sigma T^{0}\nonumber\\[2ex]
    &+\mathrm{i} h_{\omega,k}\gamma_\mu \omega_\mu T^{0}+\mathrm{i} h_{\rho,k}\gamma_\mu \bm{\rho}_\mu\cdot \bm{T}\Big]\psi_N\nonumber\\[2ex]
    &+\frac{1}{2}\partial_\mu \sigma \partial_\mu \sigma+\frac{1}{2}m_\sigma^2 \sigma^2+\frac{\lambda_3}{3!}\sigma^3+\frac{\lambda_4}{4!}\sigma^4\nonumber\\[2ex]
    &+\frac{1}{4}\omega_{\mu \nu}\omega_{\mu \nu}+\frac{1}{2}m_\omega^2\omega_\mu \omega_\mu+\frac{1}{4}\bm{\rho}_{\mu \nu}\cdot\bm{\rho}_{\mu \nu}\nonumber\\[2ex]
    &+\frac{1}{2}m_\rho^2\bm{\rho}_\mu \cdot\bm{\rho}_\mu+V_k(\sigma, \omega, \rho)\bigg\}\,,\label{eq:action}
\end{align}
which is also known as the Walecka model \cite{Walecka:1974qa}. The subscript $k$ denotes the infrared (IR) cutoff, that is, the renormalization group (RG) scale in the fRG. The field $\Phi=(\psi_N, \bar{\psi}_N, \sigma, \omega, \rho)$ includes the nucleon field, i.e., the isospin doublet of proton and neutron $\psi_N=(\psi_p, \psi_n)^\intercal$, the isoscalar-scalar $\sigma$ meson field, the isoscalar-vector $\omega$ meson field, and the isovector-vector $\rho$ meson field. The field strength tensors of vector mesons read
\begin{align}
    \omega_{\mu \nu}=&\partial_\mu \omega_\nu-\partial_\nu \omega_\mu\,,\\[2ex]
    \bm{\rho}_{\mu \nu}=&\partial_\mu \bm{\rho}_\nu-\partial_\nu \bm{\rho}_\mu\,.\label{}
\end{align}

In \Eq{eq:action} the Euclidean spacetime has been used with $\int_x=\int_0^{1/T} \mathrm{d} x_0 \int \mathrm{d}^3 x$, where $T$ is the temperature. The $m_N$ denotes the nucleon mass in the vacuum and $\hat\mu_N=\mathrm{diag}(\mu_p,\mu_n)$ the diagonal matrix of nucleon chemical potentials. The $m_\sigma$, $m_\omega$ and $m_\rho$ stand for the meson masses in the vacuum for the respective mesons. Moreover, in the vacuum there are self-interactions of cubic and quartic terms for the $\sigma$ field with the coupling strength $\lambda_3$ and $\lambda_4$, respectively. Note that all the observables in the vacuum, i.e., $m_N$, $m_\sigma$, $m_\omega$, $m_\rho$, $\lambda_3$ and $\lambda_4$, are $k$-independent.

The nucleons in \labelcref{eq:action} interact with mesons through the Yukawa couplings with the strength $h_{\sigma,k}$, $h_{\omega,k}$ and $h_{\rho,k}$, that are RG scale $k$-dependent. The mesons are in the adjoint representation of the $\mathrm{U}(N=2)$ group in the isospin space with
\begin{align}
    T^{0}=\frac{1}{2}\mathbb{1}_{2\times 2}\,,\label{}
\end{align}
and
\begin{align}
    T^i=\frac{\sigma^i}{2} \quad(i=1,\,2,\,3)\,, \label{}
\end{align}
where $\sigma^i$ are the Pauli matrices. It is readily verified that the temporal component of vector fields should be purely imaginary resulting from the requirement of the hermitian property of the effective action. 

The last term in \labelcref{eq:action} $V_k$ is the effective potential arising from density fluctuations, and thus it is vanishing in the vacuum, i.e., 
\begin{align}
    V_{k\to \infty}=0\,, \label{eq:Vk-inf}
\end{align}
In the density-fRG, quantum fluctuations in the vacuum are separated from the density fluctuations, and only the density fluctuations are taken into account through the evolution of the RG scale $k$. The vacuum case corresponds to $k\to \infty$, and in the other infrared limit $k\to 0$ the full density effects are encoded.

\section{Flow of the effective potential}
\label{sec:potential} 

\begin{figure}[t]
\includegraphics[width=0.35\textwidth]{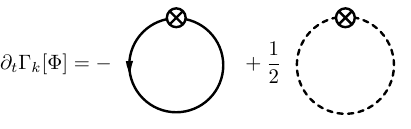}
\caption{Diagrammatic representation of the flow equation of the effective action. The solid and dashed lines stand for the nucleon and meson propagators, respectively, where all the mesons are included. The crossed circles denote the infrared regulators.}
\label{fig:action-flow}
\end{figure}

In the functional renormalization group approach, the RG scale evolution of the effective action in \Eq{eq:action} is governed by the Wetterich equation \cite{Wetterich:1992yh}, viz.,
\begin{align}
    \partial_t \Gamma_{k}[\Phi]&=-\mathrm{Tr}\Big[\big(\partial_t R_{\bar \psi \psi,k}\big) G_{\psi \bar \psi}\Big]+\frac{1}{2}\mathrm{Tr}\Big[\big(\partial_t R_{\phi \phi,k}\big) G_{\phi \phi}\Big]\,, \label{eq:WetterichEq}
\end{align}
where the terms on the right side denote the nucleon and meson loop, respectively. Here the notation $t=\ln(k/\Lambda)$ is used, which is usually named as the RG time. The $\Lambda$ is a reference scale, for instance the initial scale of the flow equation in the ultraviolet (UV). The flow equation in \Eq{eq:WetterichEq} is also diagrammatically shown in \Fig{fig:action-flow}. Note that here $\phi=(\sigma, \omega, \rho)$ includes all the mesons. The $G$ and $R$ stand for the respective propagators and regulators. In the fRG the regulator is used to suppress the fluctuations of momenta $p \lesssim k$. As a consequence, density fluctuations of different scales are successively integrated in as the flow equation in \Eq{eq:WetterichEq} is evolved from the ultraviolet $k \to \infty$ to the infrared $k \to 0$. For more details about the concept and application of the fRG method, see recent reviews, e.g., \cite{Dupuis:2020fhh, Fu:2022gou}.

%
%

Substituting the nucleon-meson effective action in \Eq{eq:action} into the flow equation in \Eq{eq:WetterichEq} and evaluating both sides on the equations of motion (EoM) for the fields $\Phi_{\mathrm{EoM}}$, i.e., the expectation values of fields in the vacuum or at a finite density, one is able to obtain the flow equation of effective potential. Note that the expectation values of fermionic fields are zero, but those of meson fields could be nonzero. Moreover, for the vector fields one has
\begin{align}
    \omega_\mu \big|_{\Phi_{\mathrm{EoM}}} =&\omega_0 \delta_{\mu 0}\,,\\[2ex]
    \rho_\mu^{i} \big|_{\Phi_{\mathrm{EoM}}} =&\rho_{03} \delta_{\mu 0}\delta_{i 3}\,,\label{}
\end{align}
which indicates that the expectation value of vector mesons is nonvanishing only when its component is the temporal one. There is another additional constraint that the isospin index should be $i=3$ for the $\rho$ meson. 

After a direct calculation, one arrives at the flow of effective potential,
\begin{align}
    \partial_t V_{k}(\sigma, \omega_0, \rho_{03})=&-\frac{2}{3 \pi^2}k^4 \Big[\Delta\mathcal{F}_{(1)}(\bar m_N^{*2};T,\mu_p^*)\nonumber\\[2ex]
    &+\Delta\mathcal{F}_{(1)}(\bar m_N^{*2};T,\mu_n^*)\Big]\,, \label{eq:dtV}
\end{align}
where $T$ stands for the temperature, and the effective chemical potentials for the proton and neutron read
\begin{align}
    \mu_p^* =&\mu_p-\frac{1}{2}\Big(h_{\omega,k}\mathrm{i} \omega_0+h_{\rho,k} \mathrm{i} \rho_{03}\Big)\,,\label{eq:mup}\\[2ex]
    \mu_n^* =&\mu_n-\frac{1}{2}\Big(h_{\omega,k}\mathrm{i} \omega_0-h_{\rho,k} \mathrm{i} \rho_{03}\Big)\,,\label{eq:mun}
\end{align}
respectively. The effective mass of nucleons is given by
\begin{align}
    m_N^*= m_N-\frac{1}{2}h_{\sigma,k}\sigma\,,\label{eq:mN}
\end{align}
which is the same for both the proton and neutron. One has $\bar m_N^*=m_N^*/k$ in \Eq{eq:dtV}. The threshold function $\mathcal{F}_{(1)}$ is defined in \Eq{eq:Fn} and its explicit expression is given in \Eq{eq:F1}. Note that in \Eq{eq:dtV} the vacuum part in the threshold function is subtracted, to wit, 
\begin{align}
    \Delta\mathcal{F}_{(1)}(\bar{m}_{f}^{2};T,\mu)= \mathcal{F}_{(1)}(\bar{m}_{f}^{2};T,\mu)-\mathcal{F}_{(1)}(\bar{m}_{f}^{2};0,0)\,,\label{eq:DeltaF1}
\end{align}
which indicates that quantum fluctuations are neglected and density fluctuations, i.e., the medium effect of high densities, are included. Note that here the temperature fluctuations are negligible, since we are interested in the case of cold nuclear matter with $T \to 0$. Since the meson propagators do not carry nucleon chemical potentials, they do not contribute to the density fluctuations. This is the reason why the meson loop in \Fig{fig:action-flow} does not contribute to the flow of effective potential in \Eq{eq:dtV}.

%
\begin{figure*}[t]
\includegraphics[width=0.8\textwidth]{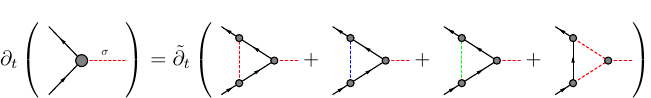}
\caption{Diagrammatic representation of the flow equation of the nucleon--$\sigma$-meson vertex. The grey blobs denote the full vertices and the lines the full propagators. The red, blue, and green dashed lines represent the $\sigma$, $\omega$, $\rho$ mesons, respectively. The derivative $\tilde{\partial}_t$ stands for that it only hits the $k$-dependence of regulators, which would result in the insertion of a regulator for each inner line of diagrams on the right side of the flow equation. Note that the last diagram on the right hand side stems from the self-interaction of the $\sigma$ meson as shown in the action \labelcref{eq:action}.}
\label{fig:nucleonsigma-equ}
\end{figure*}
%

%
\begin{figure*}[t]
\includegraphics[width=0.7\textwidth]{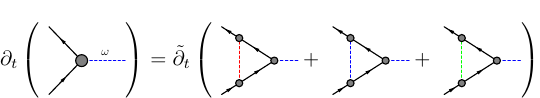}
\caption{Diagrammatic representation of the flow equation of the nucleon--$\omega$-meson vertex.}
\label{fig:nucleonomega-equ}
\end{figure*}
%

%
\begin{figure*}[t]
\includegraphics[width=0.7\textwidth]{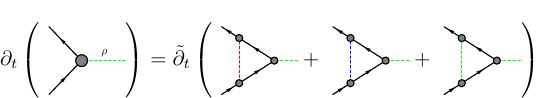}
\caption{Diagrammatic representation of the flow equation of the nucleon--$\rho$-meson vertex.}
\label{fig:nucleonrho-equ}
\end{figure*}
%

It is straightforward to integrate out the flow equation in \Eq{eq:dtV}, which yields
\begin{align}
    &V_{k=0}(\sigma, \omega_0, \rho_{03})\nonumber\\[2ex]
    =&-\frac{1}{3 \pi^2}\int_0^\infty \mathrm{d}k \frac{k^4}{E_k}\bigg[n_{F}(\bar m_N^{*2};T,\mu_p^*)+n_{F}(\bar m_N^{*2};T,-\mu_p^*)\nonumber\\[2ex]
    &+n_{F}(\bar m_N^{*2};T,\mu_n^*)+n_{F}(\bar m_N^{*2};T,-\mu_n^*)\bigg] \,,\label{eq:Vk0}
\end{align}
with
\begin{align}
    E_k=\sqrt{k^2+m_N^{*2}}\,,\label{eq:Ek}
\end{align}
and the fermionic distribution function 
\begin{align}
    n_{F}(\bar{m}_{f}^{2};T,\mu)=\frac{1}{\exp\bigg\{\frac{1}{T}\Big[k(1+\bar{m}_{f}^{2})^{1/2}-\mu\Big]\bigg\}+1}\,.\label{eq:nf}
\end{align}
Note that the initial condition for the effective potential in the ultraviolet limit $k \to \infty$ in \Eq{eq:Vk-inf} has been used. Then, one arrives at the thermodynamic potential density,
\begin{align}
    \Omega=&V_{k=0}+\frac{1}{2}m_\sigma^2 \sigma^2+\frac{\lambda_3}{3!}\sigma^3+\frac{\lambda_4}{4!}\sigma^4\nonumber\\[2ex]
    &-\frac{1}{2}m_\omega^2 (\mathrm{i}\omega_0)^2 -\frac{1}{2}m_\rho^2(\mathrm{i} \rho_{03})^2\,,\label{eq:Omega}
\end{align}
where $\mathrm{i}\omega_0$ and $\mathrm{i} \rho_{03}$ are real-valued, as discussed above. The expectation values of the field $\sigma$, $\omega_0$ and $\rho_{03}$ are determined by their respective equation of motion (EoM), i.e., the stationary conditions, 
\begin{align}
    \frac{\partial \Omega}{\partial \sigma}=\frac{\partial \Omega}{\partial \omega_0}=\frac{\partial \Omega}{\partial \rho_{03}}=0\,.\label{eq:EoM}
\end{align}

Note that if the $k$-dependence of the Yukawa couplings in \Cref{eq:mup,eq:mun,eq:mN} is neglected and their values are set to be constants, in another word, the Yukawa couplings do not depend on the density, the effective potential in \Eq{eq:dtV} or \Eq{eq:Vk0} would return to the result obtained in the mean field approximation, which is usually referred to as the relativistic mean field (RMF) calculation in the literature, see, e.g., \cite{Fu:2008zzg, Fu:2008bu, Bai:2021wrh} for more details. Our interest is obvious beyond the RMF in this work. In the next section, we will investigate the RG scale dependence of the Yukawa couplings by means of their flow equations, through which the density dependence is encoded self-consistently.

\section{Density dependent Yukawa couplings}
\label{sec:Yukawa} 

As we have discussed above, in the RMF calculations the coupling strengths are constant, independent of densities. The coupling strengths are usually fixed by fitting observables at the saturation density of the nuclear matter $\rho_0\approx 0.16 \,\mathrm{fm}^{-3}$. This is reasonable if we are only interested in the physics not far away from the saturation density. Nevertheless, if the high density physics is concerned, e.g., those involved in heavy-ion collisions and neutron star where the baryon density can be increased up to two or even several times $\rho_0$, a constant coupling strength is obviously insufficient.

In this work, we would like to use the fRG to include the contribution of density fluctuations to the running of Yukawa couplings. Since the running of Yukawa couplings varies at different baryon densities, the density dependence of interaction strengths is included self-consistently via the flows of the Yukawa couplings.

%
\begin{figure}[t]
\includegraphics[width=0.4\textwidth]{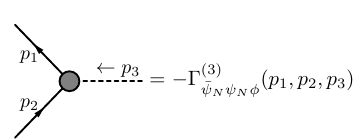}
\caption{Diagrammatic representation of the nucleon-meson vertex.}
\label{fig:nucleonmeson3}
\end{figure}
%

The flow equations of the Yukawa couplings can be obtained from the flows of the nucleon-meson vertices as shown in \Cref{fig:nucleonsigma-equ,fig:nucleonomega-equ,fig:nucleonrho-equ}. The flow of nucleon-sigma coupling reads
\begin{align}
    &\partial_t h_{\sigma,k}\nonumber\\[2ex]
    =&-\frac{1}{2}\mathrm{Re}\Big[\mathrm{Tr}\Big(T^0\partial_t \Gamma^{(3)}_{\bar \psi \psi \sigma}(p_1=p_0,p_2=p_0,p_3=0)\Big)\Big]\,,\label{eq:dth-sigma}
\end{align}
where the momentum labels are indicated in \Fig{fig:nucleonmeson3}. Here, the nucleon-meson vertex is projected onto the momenta configuration, that the meson momentum is vanishing and the fermion momentum is $p_1=p_2=p_0$ with
\begin{align}
   p_0=(p_0^0=\pi T,\, \bm{p}_0=0)\,,\label{eq:p0}
\end{align}
where the spatial component is zero and the temporal component is located at the lowest-order Matsubara frequency. In \Eq{eq:dth-sigma} the trace operates both in the Dirac and isospin spaces. Note that the expression in the square bracket is usually complex-valued at finite temperature and chemical potential, and choosing its real part is equivalent to averaging this expression with $p_0$ and $-p_0$. Thus, this expression is real as $T \to 0$.

The flow of the nucleon-omega coupling is obtained by making the projection as follows
\begin{align}
    \partial_t h_{\omega,k}=-\frac{1}{6}\mathrm{Re}\Big[\mathrm{Tr}\Big(T^0 \mathrm{i} \vec{\gamma}_\mu\partial_t  \big(\Gamma^{(3)}_{\bar \psi \psi \omega}\big)_\mu(p_0)\Big)\Big]\,,\label{eq:dth-omega}
\end{align}
where $\vec{\gamma}_\mu$ is the spatial component of the $\gamma$ matrix, that is
\begin{align}
   \vec{\gamma}_\mu=(1-\delta_{\mu 0})\gamma_\mu\,.\label{}
\end{align}
In \Eq{eq:dth-omega} we have projected the vector vertex onto the spatial components. In fact, the Lorentz symmetry is broken at finite temperature and chemical potential. Hence, in principle it is more reasonable to distinguish the temporal and spatial Yukawa couplings for the vector vertices. In this work, we simplify the calculations by using the spatial component and neglecting its difference with respect to the temporal one. This truncation is found to be an appropriate approximation and thus commonly used in literature, see e.g., \cite{Yin:2019ebz, Fu:2019hdw}. The flow of the nucleon-rho coupling reads
\begin{align}
    \partial_t h_{\rho,k}=-\frac{1}{6}\mathrm{Re}\Big[\mathrm{Tr}\Big(T^3 \mathrm{i} \vec{\gamma}_\mu\partial_t  \big(\Gamma^{(3)}_{\bar \psi \psi \rho}\big)_\mu^{i=3}(p_0)\Big)\Big]\,,\label{eq:dth-rho}
\end{align}
where the projection is performed upon the isospin generator $T^3$.

The explicit expressions of the flows of Yukawa couplings in \Cref{eq:dth-sigma,eq:dth-omega,eq:dth-rho} are presented in \Cref{app:Yukawa}, where it can be found that the flows could be expressed in terms of a variety of threshold functions. The threshold functions usually take the form, e.g., $\mathcal{FB}_{(n_f,n_b)}$, where the subscripts correspond to the numbers of the fermionic and bosonic propagators involved in the threshold function. Similar with the threshold function $\mathcal{F}_{(1)}$ in \Eq{eq:DeltaF1}, it is necessary to separate the vacuum part from the finite density part in $\mathcal{FB}_{(n_f,n_b)}$, since in the DFT one is allowed to investigate the density effect exclusively. The medium part of threshold function is obtained as
\begin{align}
   &\Delta\mathcal{FB}_{(n_f,n_b)}(\bar{m}_{f}^{2}, \bar{m}_{b}^{2};T,\mu, p_0)\nonumber\\[2ex]
    =&\mathcal{FB}_{(n_f,n_b)}(\bar{m}_{f}^{2}, \bar{m}_{b}^{2};T,\mu, p_0)\nonumber\\[2ex]
    &-\mathcal{FB}_{(n_f,n_b)}(\bar{m}_{f}^{2}, \bar{m}_{b}^{2};0,0,0)\,.\label{eq:DeltaFB}
\end{align}
Note that at vanishing temperature and finite chemical potential there is an important property known as the Silver Blaze property, which dictates that the system can reach a finite density only when the chemical potential exceeds a critical value $\mu_c$ \cite{Cohen:2003kd}. The Silver Blaze property is broken in \Eq{eq:DeltaFB}, since the Matsubara frequency of the external fermion line in the nucleon-meson vertices is truncated at some value, such as that in \Eq{eq:p0}, rather than summed for all its values \cite{Fu:2015naa, Fu:2016tey, Fu:2019hdw}. In order to recover the Silver Blaze property, we modify \Eq{eq:DeltaFB} as such
\begin{align}
   &\Delta\mathcal{FB}_{(n_f,n_b)}(\bar{m}_{f}^{2}, \bar{m}_{b}^{2};T,\mu, p_0)\nonumber\\[2ex]
    =&\Big[\mathcal{FB}_{(n_f,n_b)}(\bar{m}_{f}^{2}, \bar{m}_{b}^{2};T,\mu, p_0)\nonumber\\[2ex]
    &-\mathcal{FB}_{(n_f,n_b)}(\bar{m}_{f}^{2}, \bar{m}_{b}^{2};0,0,0)\Big]\Big[n_{F}(\bar{m}_{f}^{2};T,\mu)\nonumber\\[2ex]
    &+n_{F}(\bar{m}_{f}^{2};T,-\mu)\Big]\,,\label{eq:DeltaFB-modi}
\end{align}
where a factor $n_{F}(\bar{m}_{f}^{2};T,\mu)+n_{F}(\bar{m}_{f}^{2};T,-\mu)$ comprised of the distribution functions for the fermion and anti-fermion is multiplied with the original expression in \Eq{eq:DeltaFB}. Note that this modification is motivated by the two-loop calculations in fRG \cite{Fu:2016tey}, where it is found that this factor always emerges after the summation for the Matsubara frequency of the external fermion line is done. This can be understood readily, since the factor has the following property at $T=0$: the factor is vanishing once the energy of fermion is above the Fermi surface and it is unity when it is below the Fermi surface. In another word, this factor would not change the value of \Eq{eq:DeltaFB} below the Fermi surface while would suppress it dramatically above the Fermi surface, that is consistent with the expectation of density flows.

\section{Numerical results}
\label{sec:num} 

%
\begin{table*}[t]
  \begin{center}
 \begin{tabular}{ccccccc}
    \hline\hline & & & & & &\\[-2ex]   
    \hspace{1.5cm}  & $h_{\sigma,k\to\infty}$ & \hspace{0.5cm} $h_{\omega,\infty}$ \hspace{0.5cm} &  $h_{\rho,\infty}$ & \hspace{0.5cm}  $m_\sigma$ [MeV]   & \hspace{0.5cm}  $\lambda_3$ [MeV] & \hspace{0.5cm}  $\lambda_4$ \\[1ex]
    \hline & & & & & &\\[-2ex]
  fRG (Set I)   & 14.4466  &16.2120 & 13.6962  & 500 & 4052.05& -127.75\\[1ex]
  fRG (Set II)   & 17.1942  &15.9231 & 13.6962  & 600 & 7385.95& -264.90\\[1ex]
  RMF   & 16.0019  &18.0813 & 8.0962  & 500 & 7052.06& -127.75\\[1ex]
    \hline\hline
  \end{tabular}
  \caption{Parameters in the nucleon-meson effective field theory within the functional renormalization group (fRG) approach. Two sets of parameters for the fRG are obtained, corresponding to $m_\sigma=500$, 600 MeV, respectively. Parameters in the relativistic mean field (RMF) approximation are also presented for comparison. Note that in the RMF approximation the Yukawa couplings are constant, independent of the RG scale $k$.} 
  \label{tab:param}
  \end{center}\vspace{-0.5cm}
\end{table*}
%

%
\begin{figure*}[t]
\includegraphics[width=0.45\textwidth]{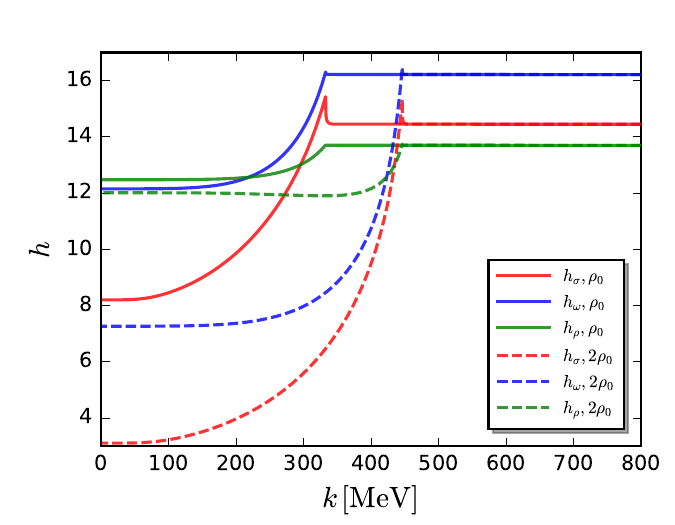}\hspace{0.3cm}
\includegraphics[width=0.45\textwidth]{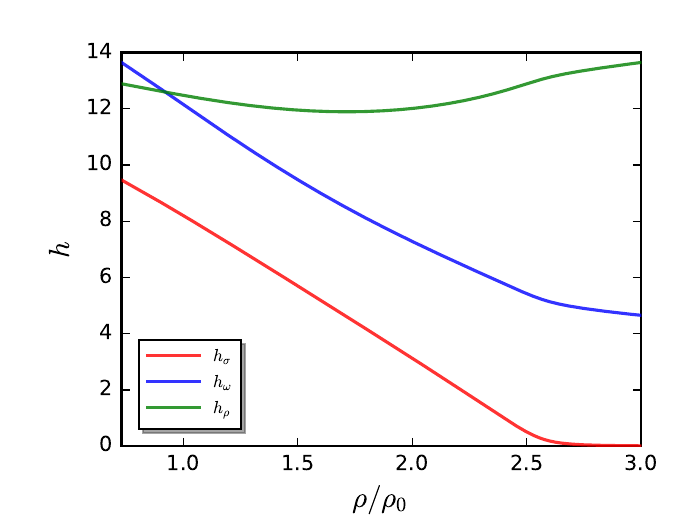}
\caption{Left panel: Yukawa couplings as functions of the RG scale $k$ at two different baryon densities, the saturation density $\rho_0$ and twice the saturation density $2\rho_0$ in the symmetric nuclear matter. Right panel: Yukawa couplings in the infrared limit $h_{k=0}$ as functions of the baryon density. Note that $m_\sigma=500$ MeV is adopted in the fRG calculations unless stated explicitly.}\label{fig:h_k}
\end{figure*}
%

To proceed, one has to determine the parameters in the nucleon-meson effective field theory in \Eq{eq:action}. The masses of vector mesons are fixed as $m_\omega=782$ MeV, $m_\rho=775$ MeV based on the particle data \cite{ParticleDataGroup:2024cfk}. The scalar $\sigma$ meson is usually identified as the meson $f_0(500)$, whose mass is not quite well constrained, being in the range $400\sim 550$ MeV. In this work, we choose two representative values $m_\sigma=500$, 600 MeV. The nucleon mass in the vacuum is given by $m_N=938$ MeV. The three Yukawa couplings in the UV limit $k \to \infty$, that is, in the vacuum, and the self-interaction strengths of $\sigma$ meson, $\lambda_3$ and $\lambda_4$, are determined by fitting the properties of symmetric nuclear matter at saturation density: The saturation baryon density $\rho_0=0.16\, \mathrm{fm}^{-3}$, the binding energy per nucleon $E/A=-16$ MeV, the incompressibility $K=215$ MeV, and the symmetry energy $E_s=32$ MeV. The obtained parameters are presented in \Tab{tab:param}. We also show the values of parameters in the RMF calculations for comparison, where the Yukawa couplings do not depend on the RG scale $k$. Note that in the RMF approximation, the calculated results only depend on the ratios $h_\sigma^2/m_\sigma^2$, $h_\omega^2/m_\omega^2$, $h_\rho^2/m_\rho^2$, $\lambda_3/h_\sigma^3$, $\lambda_4/h_\sigma^4$, rather than the specific values of meson masses.

In the left panel of \Fig{fig:h_k} we show the three different Yukawa couplings $h_{\sigma,k}$, $h_{\omega,k}$ and $h_{\rho,k}$ as functions of the RG scale $k$ for two different baryon densities: One is the saturation density $\rho_0$ and the other twice the saturation density $2\rho_0$. Note that parameter Set I in \Tab{tab:param} is used as the default parameters, that is, $m_\sigma=500$ MeV is adopted unless stated explicitly. One can see that when $k$ is large, there is no density fluctuation and the Yukawa couplings stay constant. When $k$ is below some threshold value, the density fluctuations start to play a role and the Yukawa couplings evolve and deviate from their vacuum values. This threshold value of $k$ for the start of the density fluctuations increases with the baryon density of nuclear matter, which can be easily found by comparing the Yukawa running at $\rho=\rho_0$ and $\rho=2\rho_0$. Moreover, it is found that as the RG scale $k$ is lowered down, the Yukawa couplings become smaller, except that $h_{\sigma,k}$ increases a bit in the early period of evolution. This can be interpreted as that the interaction strength between the nucleons and mesons is screened by the nuclear medium. In the right panel of \Fig{fig:h_k} the Yukawa couplings in the IR limit, i.e., those at $k=0$, are shown as functions of the baryon density. One finds that $h_{\sigma,k=0}$ and $h_{\omega,k=0}$ decrease with the increase of the baryon density, and when $\rho \gtrsim 2.5 \,\rho_0$ the $h_{\sigma,k=0}$ is very close to the zero. The dependence of $h_{\rho,k=0}$ on the density is relatively milder in comparison to the Yukawa couplings of $\sigma$ and $\omega$ mesons.

%
\begin{figure*}[t]
\includegraphics[width=0.45\textwidth]{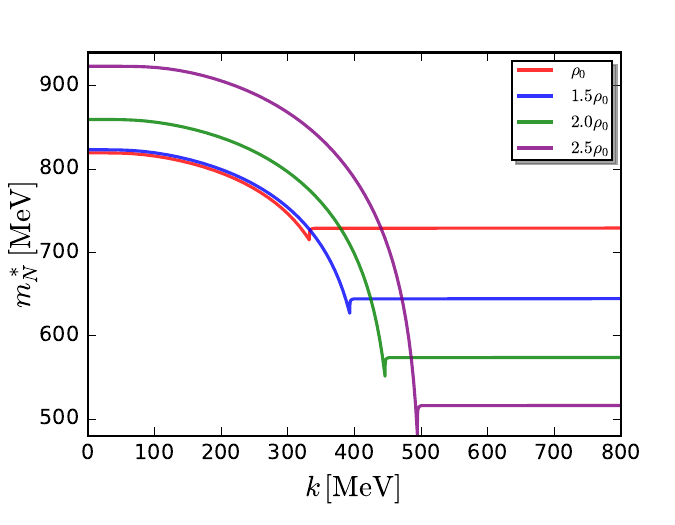}\hspace{0.3cm}
\includegraphics[width=0.45\textwidth]{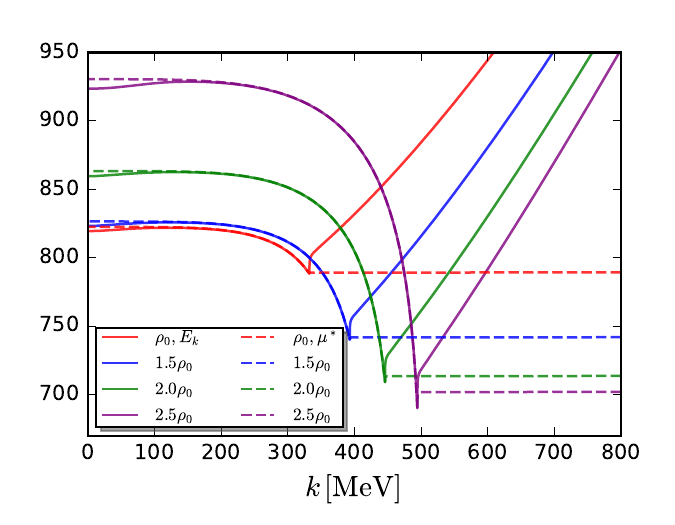}
\caption{Effective nucleon mass $m_N^*= m_N-(h_{\sigma,k}\sigma)/2$ (left panel) and effective chemical potential $\mu^* =\mu-(h_{\omega,k}\mathrm{i} \omega_0)/2$ (right panel) as functions of the RG scale $k$ at several different baryon densities in the symmetric nuclear matter. The effective chemical potential is also compared with the effective nucleon energy $E_k=\sqrt{k^2+m_N^{*2}}$.}\label{fig:mass-chemi-pot}
\end{figure*}
%

%
\begin{figure}[t]
\includegraphics[width=0.45\textwidth]{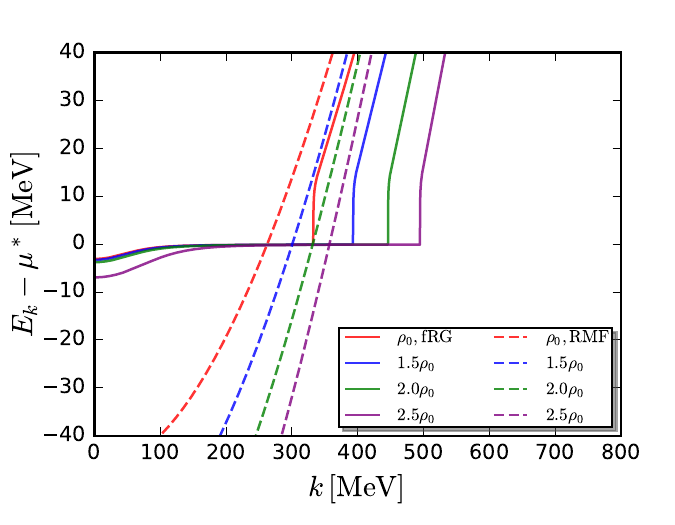}
\caption{Difference between the effective energy and the effective chemical potential $E_k-\mu^*$ as a function of the RG scale $k$ at several different baryon densities in the symmetric nuclear matter, where the fRG and RMF are compared.}
\label{fig:Diff-Ek-mu}
\end{figure}
%

%
\begin{figure}[t]
\includegraphics[width=0.45\textwidth]{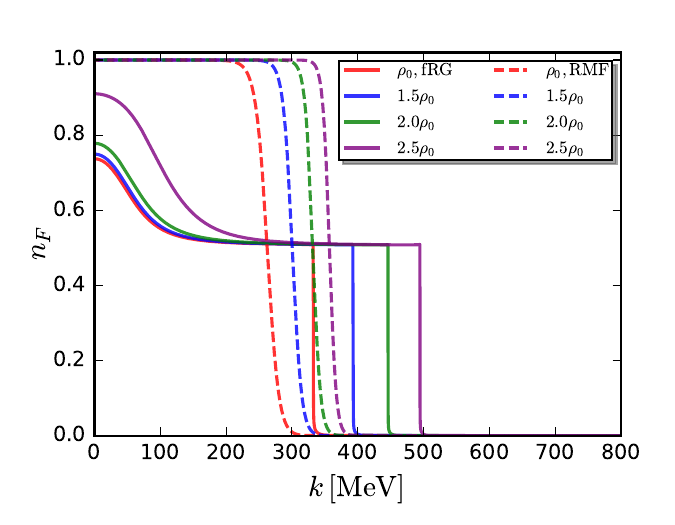}
\caption{Fermionic distribution function $n_{F}$ as a function of the RG scale $k$ at several different baryon densities in the symmetric nuclear matter.}
\label{fig:nf}
\end{figure}
%

%
\begin{figure*}[t]
\includegraphics[width=0.45\textwidth]{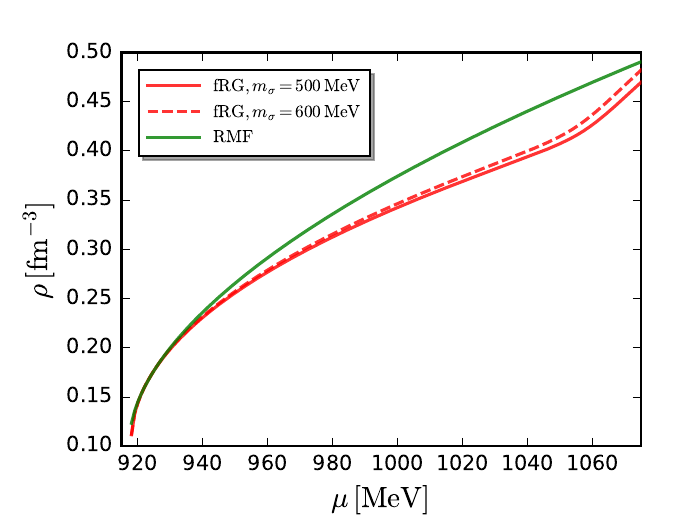}\hspace{0.3cm}
\includegraphics[width=0.45\textwidth]{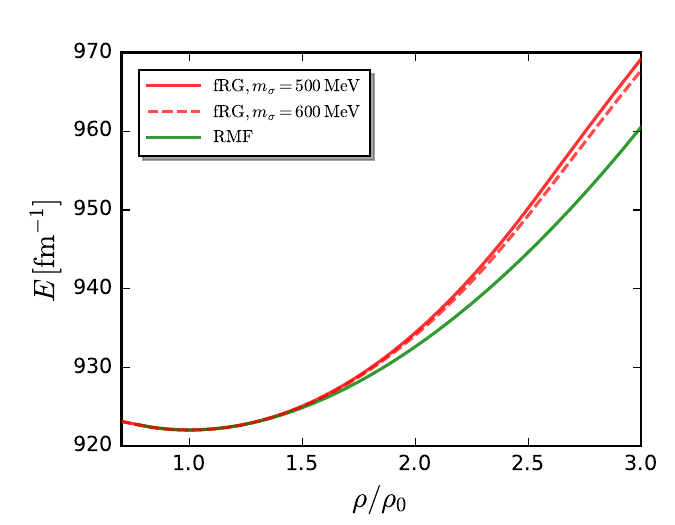}
\caption{Left panel: Nucleon density as a function of the nucleon chemical potential $\mu=\mu_p=\mu_n$ in the symmetric nuclear matter, where the RMF and fRG with $m_\sigma=500$, 600 MeV are compared. Right panel: Energy per nucleon as a function of the nucleon density.}
\label{fig:rho-E}
\end{figure*}
%

%
\begin{figure*}[t]
\includegraphics[width=0.45\textwidth]{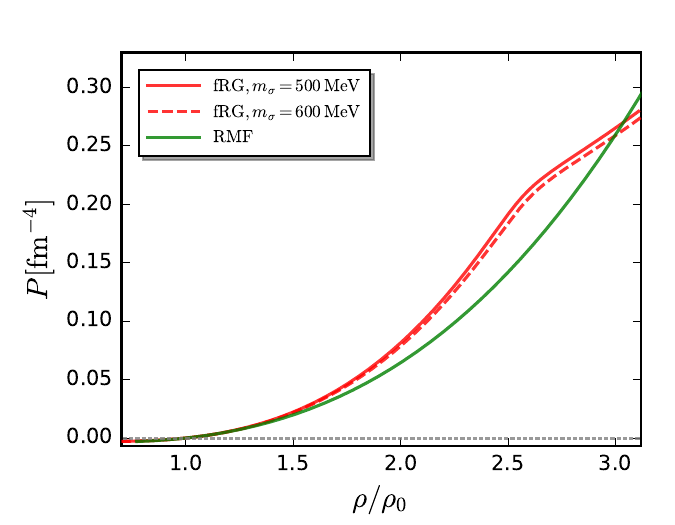}\hspace{0.3cm}
\includegraphics[width=0.45\textwidth]{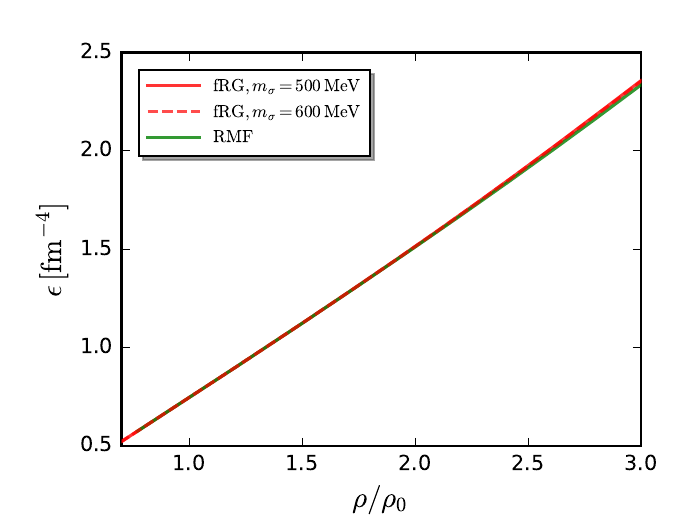}
\caption{Pressure (left panel) and energy density (right panel) as functions of the nucleon density in the symmetric nuclear matter, where the RMF and fRG with $m_\sigma=500$, 600 MeV are compared.}
\label{fig:pre_ener_dens}
\end{figure*}
%

We would like to pause here and discuss the reliability of the nucleon-meson effective field theory in \Eq{eq:action} in the regime of high baryon densities. For a crude estimate it is convenient to define a ratio as follows
\begin{align}
   R=\frac{2 r_N}{\rho^{-1/3}}\,,\label{eq:R-overlap}
\end{align}
where $r_N$ denotes the radius of nucleon, and thus the numerator represents the diameter of nucleon. Here we adopt the value of proton charge radius for $r_N$, which was found to be $r_N=0.84\sim 0.88$ fm through experimental measurement \cite{Lee:2015jqa, Bezginov:2019mdi}. The denominator in \Eq{eq:R-overlap} represents the average distance between two nucleons. Then, one immediately arrives at $R=0.91\sim 0.96$ for $\rho=\rho_0$, $R=1.04\sim 1.09$ for $\rho=1.5\rho_0$, $R=1.15\sim 1.20$ for $\rho=2\rho_0$, $R=1.24\sim 1.30$ for $\rho=2.5\rho_0$, and $R=1.32\sim 1.38$ for $\rho=3\rho_0$. It is remarkable to note that when the baryon density is just about 1.5 times the saturation density, the nucleons begin to overlap with each other. The overlap factor in \Eq{eq:R-overlap} grows up to $\sim 1.3$ when the density is increased to $\rho=2.5\rho_0$, where the substructure of nucleon and the resulting dynamics should begin to play a role, that are missing in the nucleon-meson effective theory. In short, the reliability of calculations in this work is reasonably doubted in the high density regime of, e.g., $\rho \gtrsim 3\rho_0$.

The running Yukawa couplings lead directly to the running effective nucleon mass and effective chemical potentials as well, as shown in \Eq{eq:mN,eq:mup,eq:mun}, in sharp contrast to the RMF case, where both the effective mass and the effective potential are constants at a fixed value of baryon density. It is expectable that these new properties would give rise to novel effects. In \Fig{fig:mass-chemi-pot} the effective mass and effective chemical potential are depicted as functions of the RG scale $k$. The expectation value of $\rho$ meson is vanishing in the symmetric nuclear matter. It is found from the left panel of \Fig{fig:mass-chemi-pot} that the effective nucleon mass develops a little dip and increases subsequently, after the RG scale $k$ is lowered down below the threshold value. Moreover, with the increase of the baryon density the effective mass is smaller in the UV, i.e., in the region of large $k$, which is similar with the RMF calculation. However, the effective mass increases more significantly for larger $\rho$ and its value is larger in the IR, which is quite different from the RMF. One can see that when the baryon density is increased up to $\rho=2.5\rho_0$, the effective nucleon mass at $k=0$ is close to the nucleon mass in the vacuum.

In the right panel of \Fig{fig:mass-chemi-pot} one finds that the effective chemical potential increases as well when the RG scale $k$ is below some threshold value. Furthermore, we also compare the energy of quasi-nucleon in \Eq{eq:Ek}, named the effective nucleon energy hereafter, with the effective chemical potential. It is found that in the region of large $k$ the effective energy is larger than the effective chemical potential and there are no flows or particles. When the RG scale $k$ is below the threshold value, the effective energy is smaller than the effective chemical potential, but in the subsequent sizable region of $k$, the effective energy is very close to the effective chemical potential. This is a novel phenomenon and has never been observed in the RMF calculations, see \Fig{fig:Diff-Ek-mu}, where $E_k-\mu^*$ is shown, and the fRG and RMF are compared. This phenomenon is not accidental since it takes place for all the results with different baryon densities as shown in the right panel of \Fig{fig:mass-chemi-pot} and \Fig{fig:Diff-Ek-mu}.  

It is interesting to investigate how this phenomenon takes place. Notice that the running of the Yukawa couplings is possible only when the effective energy is below the effective chemical potential, i.e., below the Fermi surface. In another word, the flows of Yukawa couplings are driven only when the solid line is below the dashed line in the right panel of \Fig{fig:mass-chemi-pot}. One can see that the Fermi surface is also running with the RG scale $k$. Once the solid line is below the dashed line, the screening effect begins to play a role and it increases the effective mass, so the solid line is pushed toward the dashed line, but cannot be pushed over the dashed line, otherwise, it will cease the running of flows. Consequently, the solid and dashed lines are inclined to stay together. We would like to call this phenomenon the locking between the effective energy and the effective chemical potential, or the locking of Fermi surface concisely.

Further investigation reveals that the vector exchange in the flow of $h_{\sigma,k}$, that is, the second and third diagrams in \Fig{fig:nucleonsigma-equ}, plays a pivotal role in the screening effect of nuclear medium in the attractive $\sigma$ channel. In \Fig{fig:nf} we show the fermionic distribution function in \Eq{eq:nf} as a function of the RG scale $k$. Due to the locking of Fermi surface, the distribution function is not far away from 1/2 in a sizable region of $k$ below the threshold value. In the actual calculations the temperature is chosen to be $T=3$ MeV, and even lower temperature makes the numerical calculations very difficult.

Solving the flow equations of the effective potential and the Yukawa couplings, discussed in \Cref{sec:potential} and \Cref{sec:Yukawa} respectively, alongside the EoM in \Eq{eq:EoM}, one is able to obtain the thermodynamic potential density in \Eq{eq:Omega}. Then the pressure reads
\begin{align}
   p=-\Omega(T, \mu_p, \mu_n)\,.\label{}
\end{align}
The proton and neutron densities are given by
\begin{align}
   \rho_p=-\frac{\partial \Omega}{\partial \mu_p},\qquad \rho_n=-\frac{\partial \Omega}{\partial \mu_n}\,,\label{}
\end{align}
respectively, and the nucleon density thus reads
\begin{align}
   \rho=\rho_p+\rho_n\,.\label{}
\end{align}
The entropy density is given by
\begin{align}
   s=-\frac{\partial \Omega}{\partial T}\,.\label{}
\end{align}
Then one arrives at the energy density 
\begin{align}
   \varepsilon=Ts+\mu_p \rho_p+\mu_n \rho_n-p\,,\label{}
\end{align}
and the energy per nucleon
\begin{align}
   E=\frac{\varepsilon}{\rho}\,.\label{}
\end{align}

The calculated density and the energy per nucleon are presented in the left and right panel of \Fig{fig:rho-E}, respectively. In \Fig{fig:pre_ener_dens} we show the pressure and energy density as functions of the nucleon density. The fRG results are compared with those of the RMF. Here, both parameter Set I and Set II in \Tab{tab:param} are employed, that is, two different sigma masses $m_\sigma=500$, 600 MeV are chosen in the fRG calculations. It is found that the pressure in fRG increases faster than that in RMF when the density is below $\sim 2.5 \rho_0$, that is attributed to the screening effect discussed in \Fig{fig:h_k} and \Fig{fig:mass-chemi-pot}. When the density is increased up to $\rho \gtrsim 2.5 \rho_0$, the screening effect is almost finished, evidenced in the right panel of \Fig{fig:h_k}, and thus the increase of the pressure slows down. This is also reflected in the dependence of density on the nucleon chemical potential in the left panel of \Fig{fig:rho-E}. It should be cautioned that as we discussed above, the reliability of the nucleon-meson EFT used in this work is gradually lost in the region of $\rho  \gtrsim (2.5 \sim 3) \rho_0$.

\section{Interpolation between the fRG and RMF}
\label{sec:Interpolation} 

%
\begin{figure*}[t]
\includegraphics[width=0.45\textwidth]{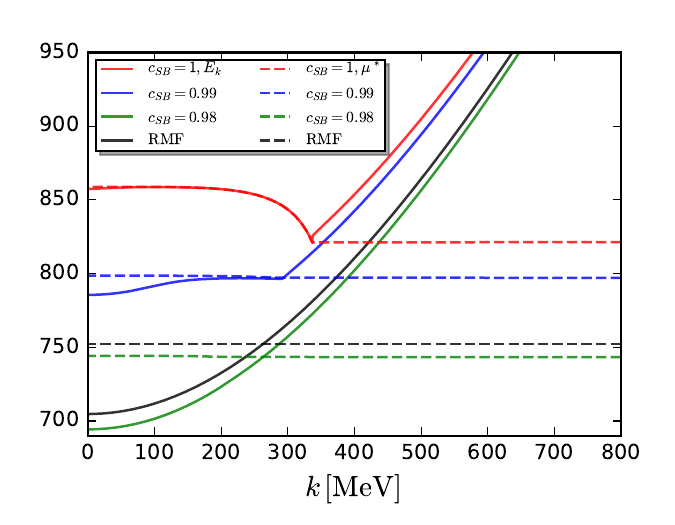}\hspace{0.3cm}
\includegraphics[width=0.45\textwidth]{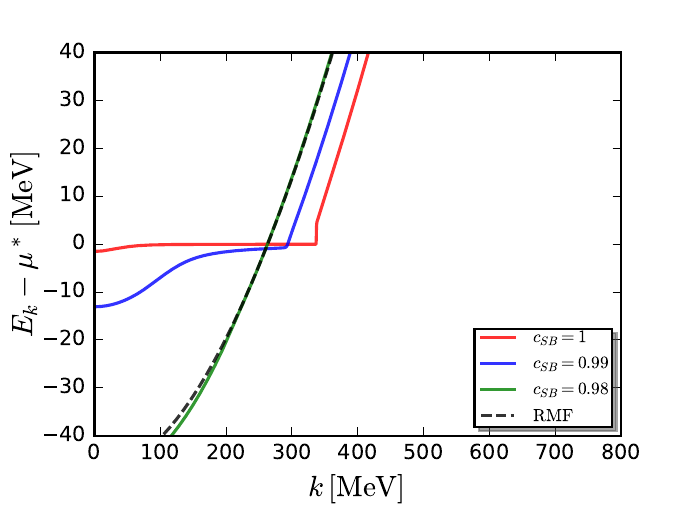}
\caption{Left panel: Effective nucleon energy $E_k$ and effective chemical potential $\mu^*$ as functions of the RG scale $k$ with $\rho=\rho_0$ in the symmetric nuclear matter, where the fRG results with several different values of $c_{_{SB}}$ as well as the RMF results are compared. Right panel: Difference between the effective energy and the effective chemical potential $E_k-\mu^*$ as a function of the RG scale $k$ with $\rho=\rho_0$ in the symmetric nuclear matter, where the fRG results with several different values of $c_{_{SB}}$ as well as the RMF results are compared.}\label{fig:energy-chemi-pot-SB}
\end{figure*}
%

%
\begin{figure}[t]
\includegraphics[width=0.45\textwidth]{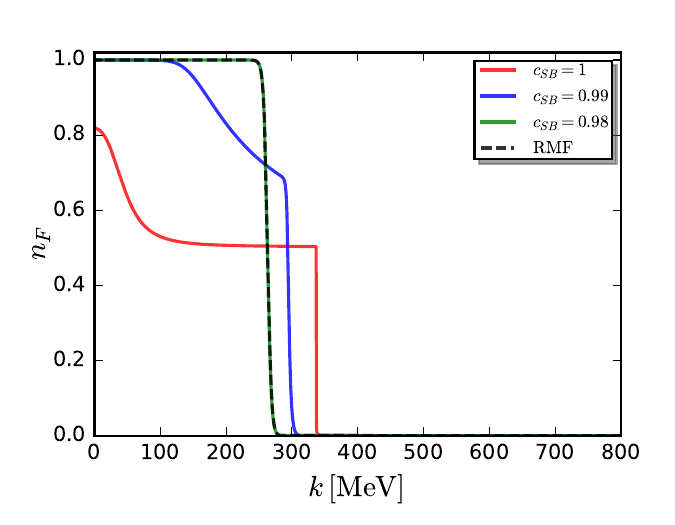}
\caption{Fermionic distribution function $n_{F}$ as a function of the RG scale $k$ with $\rho=\rho_0$ in the symmetric nuclear matter, where the fRG results with several different values of $c_{_{SB}}$ as well as the RMF results are compared.}
\label{fig:nf-SB}
\end{figure}
%

In \sec{sec:Yukawa} we have discussed the density flows of the Yukawa couplings. In order to recover, at least partially,  the Silver Blaze property violated by the truncation of flows, we have added a factor $n_{F}(\bar{m}_{f}^{2};T,\mu)+n_{F}(\bar{m}_{f}^{2};T,-\mu)$ in the threshold functions in \Eq{eq:DeltaFB-modi}. In fact, one can utilize the factor to gradually turn up or turn down the flows of Yukawa couplings, and interpolate between the case of fRG  and that of RMF discussed above continuously. To that end, we modify \Eq{eq:DeltaFB-modi} as such
\begin{align}
   &\Delta\mathcal{FB}_{(n_f,n_b)}(\bar{m}_{f}^{2}, \bar{m}_{b}^{2};T,\mu, p_0)\nonumber\\[2ex]
    =&\Big[\mathcal{FB}_{(n_f,n_b)}(\bar{m}_{f}^{2}, \bar{m}_{b}^{2};T,\mu, p_0)\nonumber\\[2ex]
    &-\mathcal{FB}_{(n_f,n_b)}(\bar{m}_{f}^{2}, \bar{m}_{b}^{2};0,0,0)\Big]\Big[n_{F}(\bar{m}_{f}^{2};T,c_{_{SB}}\mu)\nonumber\\[2ex]
    &+n_{F}(\bar{m}_{f}^{2};T,-c_{_{SB}}\mu)\Big]\,.\label{eq:DeltaFB-modi-cSB}
\end{align}
One immediately finds for the factor 
\begin{align}
   &\Big[n_{F}(\bar{m}_{f}^{2};T,\mu)+n_{F}(\bar{m}_{f}^{2};T,-\mu)\Big]\nonumber\\[2ex]
   \longrightarrow &\Big[n_{F}(\bar{m}_{f}^{2};T,c_{_{SB}}\mu)+n_{F}(\bar{m}_{f}^{2};T,-c_{_{SB}}\mu)\Big]\,,\label{}
\end{align}
where the nucleon chemical potential is multiplied with a constant $c_{_{SB}}$, i.e., $\mu\to c_{_{SB}}\mu$. Obviously, the flows of Yukawa couplings are turned down continuously as the constant $c_{_{SB}}$ decreases starting from $c_{_{SB}}=1$, and consequently the fRG results evolve towards the RMF results.

In \Fig{fig:energy-chemi-pot-SB} we show the effective nucleon energy, the effective chemical potential and their difference as functions of the RG scale $k$ with $\rho=\rho_0$ in the symmetric nuclear matter, obtained with three different values of $c_{_{SB}}$, i.e., $c_{_{SB}}=0.98$, 0.99, 1, which are also compared with the RMF results. The relevant results of the fermionic distribution function are presented in \Fig{fig:nf-SB}. It is found that when the value of the constant $c_{_{SB}}$ is reduced to 0.98, the results of fRG are very close to the RMF, and those of $c_{_{SB}}=0.99$ lie in between, which is consistent with the expectations.

\section{Conclusions and outlook}
\label{sec:conclusion}

In this work, we have proposed the method of density functional renormalization group (density-fRG) to compute the density fluctuations within the functional renormalization group approach. The density-fRG is feasible if the density fluctuations can be separated from the vacuum, i.e., quantum fluctuations. Therefore, one is allowed to deal with the medium effect exclusively and the density fluctuations are taken into account successively through the RG evolution of flow equations, which simplifies the calculations significantly at high densities. The separation of the density and quantum fluctuations is guaranteed by the fact that they dominate in different regimes of RG scale: The density fluctuation plays a role mainly in the regime of $k \lesssim 400\sim 500 $ MeV, as shown in \Fig{fig:mass-chemi-pot} and \Fig{fig:Diff-Ek-mu}, while the quantum fluctuation usually dominates above this scale, see the first-principles QCD calculations in, e.g., \cite{Fu:2019hdw}.

The density-fRG is applied to the nucleon-meson effective field theory, also known as the Walecka model, to study the properties of nuclear matter at high baryon densities. We investigate the running of the Yukawa couplings driven by the density fluctuations in detail. It is found that the medium effect screens both the attractive nucleon-scalar-meson ($N-\sigma$) and the repulsive nucleon-vector-meson ($N-\omega$) interactions. As a consequence of the screening effect and the running Yukawa couplings, a new phenomenon called the locking of Fermi surface is found. In the locking of Fermi surface the effective energy of quasi-nucleon is always close to the Fermi surface, which are both running with the RG scale. Furthermore, the screening effect also leads to that the equation of state (EoS) of nuclear matter is stiffer in the regime of $\rho_0 \lesssim \rho \lesssim 2.5 \rho_0$ and then becomes softer when $\rho \gtrsim 2.5 \rho_0$. Note however that the reliability of the nucleon-meson EFT is gradually lost in the region of $\rho  \gtrsim (2.5 \sim 3) \rho_0$.

It is straightforward to extend the studies in this work. The investigation of the symmetry energy of nuclear matter at high densities within the density-fRG is underway, which will be reported in the near future \cite{Chen:2025a}. Moreover, in the density-fRG it is readily to include contributions from the pion meson, which is vanishing in the mean field approximation, but its fluctuations are sizable because the pion mass is significantly smaller than the masses of other mesons. Furthermore, it is also convenient to extend the nucleon-meson effective action in \Eq{eq:action} to include the degrees of freedom of hyperons. The interaction of hyperons and its modification in the high density medium are subjects of high interest and importance in nuclear physics and nuclear astrophysics. Most importantly, it is desirable to combine the theoretical results obtained in the density-fRG or even the density-fRG itself with some phenomenological transport model, such that one is able to compare with experimental data of heavy-ion collisions in the upcoming new facilities of high densities.

\section*{Acknowledgements}

We thank Jan M. Pawlowski for discussions and comments. We also would like to thank the members of the fQCD collaboration \cite{fQCD} for collaborations on related projects. This work is supported by the National Natural Science Foundation of China under Contract Nos.\ 12447102, 12175030. Y.R. Chen is supported by Alexander v. Humboldt Foundation through Humboldt Research Fellowship for postdocs.

\appendix 

\section{Some notations}
\label{app:notat}

\subsection{Propagators and infrared regulators}
\label{appsubsec:prop}

The two-nucleon correlation function reads
\begin{align}
    \Gamma^{(2)}_{\bar \psi \psi}(p^\prime,p)=\frac{\overrightarrow\delta}{\delta \bar{\psi}_N(p^\prime)} \Gamma_k[\Phi]\frac{\overleftarrow \delta}{\delta \psi_N(p)}\bigg|_{\Phi=\Phi_{\mathrm{EoM}}}\,,\label{eq:Gamma2bNN}
\end{align}
where the left and right derivatives are used for the fermionic field of nucleons. The $p$ and $p^\prime$ stand for the momenta of anti-nucleon and nucleon, respectively, and their directions are assumed to be along the direction of the fermionic flow, otherwise stated explicitly. Inserting \Eq{eq:action} into \Eq{eq:Gamma2bNN}, one arrives at
\begin{align}
    \Gamma^{(2)}_{\bar \psi \psi}(p^\prime,p)= \Gamma^{(2)}_{\bar \psi \psi}(p)(2\pi)^4\delta^4(p^\prime-p)\,,\label{}
\end{align}
with
\begin{align}
    \Gamma^{(2)}_{\bar \psi \psi}(p)= \gamma_0 (\mathrm{i} p_0-\hat\mu_N^*)+\mathrm{i} \bm{\gamma} \cdot \bm{p}+m_N^*\,,\label{}
\end{align}
where the effective chemical potential is given by
\begin{align}
    \hat\mu_N^*&=\begin{pmatrix} \mu_p^* & 0 \\ 0 & \mu_n^* \end{pmatrix}\,, \label{}
\end{align}

In this work we adopt the $3d$ optimized, i.e., flat IR regulators \cite{Litim:2000ci, Litim:2001up}, which are used to suppress fluctuations of modes with momenta smaller than the RG scale, that is, $p \lesssim k$, see, e.g., \cite{Fu:2022gou} for details. The fermionic regulator reads
\begin{align}
    R_{\bar \psi \psi}(p)=\mathrm{i} \bm{\gamma} \cdot \bm{p} \,r_{F}(\bm{p}^2/k^2)\,,\label{}
\end{align}
with the shape function
\begin{align}
    r_{F}(x)=\left(\frac{1}{\sqrt{x}}-1\right)\Theta(1-x)\,,\label{eq:rFopt}
\end{align}
where $\Theta(x)$ is the Heaviside step function. Then one arrives at the nucleon propagator with an IR regulator, as follows
\begin{align}
    G_{\psi \bar \psi}(p)=& \Big[\Gamma^{(2)}_{\bar \psi \psi}(p)+R_{\bar \psi \psi}(p)\Big]^{-1}\nonumber\\[2ex]
    =&\begin{pmatrix} G_{p}(p) & 0 \\ 0 & G_{n}(p) \end{pmatrix}\,,\label{}
\end{align}
where the proton and neutron propagator read
\begin{align}
    &G_{p/n}(p)\nonumber\\[2ex]
    =& \frac{1}{\mathrm{i} \gamma_0 (p_0+\mathrm{i} \mu_{p/n}^*)+\mathrm{i} \bm{\gamma} \cdot \bm{p}\big(1+r_{F}(\bm{p}^2/k^2)\big)+m_N^*}\,.\label{}
\end{align}

The two-point correlation function for the $\sigma$ meson reads
\begin{align}
    \Gamma^{(2)}_{\sigma \sigma}(p^\prime,p)= &\frac{\delta^2 \Gamma_k[\Phi]}{\delta \sigma(p^\prime)\delta \sigma(p)}\bigg|_{\Phi=0}\nonumber\\[2ex]
    =&\Gamma^{(2)}_{\sigma \sigma}(p)(2\pi)^4\delta^4(p^\prime+p)\,,\label{}
\end{align}
with
\begin{align}
    \Gamma^{(2)}_{\sigma \sigma}(p)=p^2+m_\sigma^2 \,.\label{}
\end{align}
Note that in this work we have not taken into account the density effect on the meson masses, and $m_\sigma$ in the equation above denotes the $\sigma$ mass in the vacuum. This also applies to the $\omega$ and $\rho$ mesons. This is because the influence of the meson masses is indirect in comparison to the Yukawa couplings. The IR regulator for the $\sigma$ meson is given by
\begin{align}
    R_{\sigma \sigma}(p)=\bm{p}^2 \,r_{B}(\bm{p}^2/k^2)\,,\label{}
\end{align}
with the bosonic shape function
\begin{align}
    r_{B}(x)=\left(\frac{1}{x}-1\right)\Theta(1-x)\,.\label{eq:rBopt}
\end{align}
Thus, the $\sigma$ propagator reads
\begin{align}
    G_{\sigma}(p)=&\frac{1}{\Gamma^{(2)}_{\sigma \sigma}(p)+R_{\sigma \sigma}(p)}\nonumber\\[2ex]
    =&\frac{1}{p_0^2+\bm{p}^2\big(1+r_{B}(\bm{p}^2/k^2)\big)+m_\sigma^2}\,.\label{}
\end{align}

For the vector meson $\omega$ we obtain the two-point correlation function as follows
\begin{align}
    \big(\Gamma^{(2)}_{\omega \omega}\big)_{\mu \nu}(p^\prime,p)= &\frac{\delta^2 \Gamma_k[\Phi]}{\delta \omega_\mu(p^\prime)\delta \omega_\nu(p)}\bigg|_{\Phi=0}\nonumber\\[2ex]
    =&\big(\Gamma^{(2)}_{\omega \omega}\big)_{\mu \nu}(p)(2\pi)^4\delta^4(p^\prime+p)\,,\label{}
\end{align}
with
\begin{align}
    \big(\Gamma^{(2)}_{\omega \omega}\big)_{\mu \nu}(p)=\Big(p^2+m_\omega^2\Big)\Pi_{\mu\nu}^{\perp}(p)+\bigg(m_\omega^2+\frac{p^2}{\xi}\bigg)\Pi_{\mu\nu}^{\parallel}(p)\,,\label{eq:Gam2-omega2}
\end{align}
where the transverse and longitudinal polarization tensors read
\begin{align}
    \Pi_{\mu\nu}^{\perp}(p)=&\delta_{\mu\nu}-\frac{p_{\mu}p_{\nu}}{p^2}\,,\\[2ex]
    \Pi_{\mu\nu}^{\parallel}(p)=&\frac{p_{\mu}p_{\nu}}{p^2}\,,\label{}
\end{align}
respectively. Note that in \Eq{eq:Gam2-omega2} we have added the gauge fixing term with the gauge parameter $\xi$. The regulator of the $\omega$ meson reads
\begin{align}
    \big(R_{\omega \omega}\big)_{\mu \nu}(p)=\bm{p}^2 \,r_{B}(\bm{p}^2/k^2)\Pi_{\mu\nu}^{\perp}(p)\,.\label{}
\end{align}
Then, one arrives at the propagator of the $\omega$ meson
\begin{align}
    G_{\omega, \mu \nu}(p)=&\Big[\Gamma^{(2)}_{\omega \omega}(p)+R_{\omega \omega}(p)\Big]^{-1}_{\mu \nu}\nonumber\\[2ex]
    =&\frac{1}{p_0^2+\bm{p}^2\big(1+r_{B}(\bm{p}^2/k^2)\big)+m_\omega^2}\Pi_{\mu\nu}^{\perp}(p)\,,\label{}
\end{align}
where we have chosen the gauge parameter $\xi=0$. In the same way, one obtains the propagator of $\rho$ meson as follows
\begin{align}
    G^{ij}_{\rho, \mu \nu}(p)=&\frac{1}{p_0^2+\bm{p}^2\big(1+r_{B}(\bm{p}^2/k^2)\big)+m_\omega^2}\Pi_{\mu\nu}^{\perp}(p)\delta_{ij}\,.\label{}
\end{align}

\subsection{Nucleon-meson vertices}
\label{appsubsec:verti}

The three-point nucleon-meson correlation function is given by
\begin{align}
    \Gamma^{(3)}_{\bar \psi \psi \phi}(p_1,p_2,p_3)=\frac{\delta}{\delta \phi(p_3)}\frac{\overrightarrow\delta}{\delta \bar{\psi}_N(p_1)} \Gamma_k[\Phi]\frac{\overleftarrow \delta}{\delta \psi_N(p_2)}\bigg|_{\Phi=\Phi_{\mathrm{EoM}}}\,,\label{}
\end{align}
which is shown diagrammatically in \Fig{fig:nucleonmeson3}. Consequently, one finds for the nucleon-sigma vertex 
\begin{align}
    \Gamma^{(3)}_{\bar \psi \psi \sigma}(p_1,p_2,p_3)=-h_{\sigma,k} T^{0}(2\pi)^4\delta^4(p_1-p_2-p_3)\,.\label{}
\end{align}
We will not show explicitly the delta function for the four-momentum conservation in the following. The nucleon-omega vertex reads
\begin{align}
    \big(\Gamma^{(3)}_{\bar \psi \psi \omega}\big)_\mu(p_1,p_2,p_3)=\mathrm{i} h_{\omega,k}\gamma_\mu T^{0}\,.\label{}
\end{align}
and the nucleon-rho vertex
\begin{align}
    \big(\Gamma^{(3)}_{\bar \psi \psi \rho}\big)_\mu^{i}(p_1,p_2,p_3)=\mathrm{i} h_{\rho,k}\gamma_\mu T^{i}\,.\label{}
\end{align}

\section{Flows of the Yukawa couplings}
\label{app:Yukawa}

As shown in \Cref{fig:nucleonsigma-equ,fig:nucleonomega-equ,fig:nucleonrho-equ}, it is convenient to decompose the flows of Yukawa couplings into several subparts corresponding to the diagrams on the right side of the flow equations. Thus, the flow of the nucleon-$\sigma$ coupling receives contributions from the four diagrams, i.e.,
\begin{align}
    \partial_t h_{\sigma,k}=\mathrm{Flow}_{h_\sigma,1}+\mathrm{Flow}_{h_\sigma,2}+\mathrm{Flow}_{h_\sigma,3}+\mathrm{Flow}_{h_\sigma,4}\,.\label{eq:dthsigma}
\end{align}
The first part of the flow is given by
\begin{align}
    &\mathrm{Flow}_{h_\sigma,1}\nonumber\\[2ex]
    =&\frac{h_{\sigma,k}^3}{24\pi^2} \Big[\mathcal{FB}^a_{(2,1)}(\bar{m}^{*2}_N,\bar{m}_\sigma^2;T,\mu^*_p,p_0)\nonumber\\[2ex]
    &\hspace{1.cm}+\mathcal{FB}^a_{(1,2)}(\bar{m}^{*2}_N,\bar{m}_\sigma^2;T,\mu^*_p,p_0) \Big]\nonumber\\[2ex]
    &-\frac{h_{\sigma,k}^3}{12\pi^2} \bar{m}^{*2}_N \Big[\mathcal{FB}^a_{(2,2)}(\bar{m}^{*2}_N,\bar{m}_\sigma^2;T,\mu^*_p,p_0)\nonumber\\[2ex]
    &\hspace{1.7cm}+2\mathcal{FB}^a_{(3,1)}(\bar{m}^{*2}_N,\bar{m}_\sigma^2;T,\mu^*_p,p_0)\Big]\nonumber\\[2ex]
    &+(p\rightarrow n)\,,
\end{align}
where the definition and explicit expressions of the threshold functions as well as those in the following are presented in \app{app:thres}. The second part of the flow in \Eq{eq:dthsigma} reads
\begin{align}
    &\mathrm{Flow}_{h_\sigma,2}\nonumber\\[2ex]
    =&-\frac{h_{\omega,k}^2 h_{\sigma,k}}{8\pi^2} \Big[\mathcal{FB}^a_{(2,1)}(\bar{m}^{*2}_N,\bar{m}_\omega^2;T,\mu^*_p,p_0) \nonumber\\[2ex]
    &\hspace{1.5cm}+\mathcal{FB}^a_{(1,2)}(\bar{m}^{*2}_N,\bar{m}_\omega^2;T,\mu^*_p,p_0) \Big]\nonumber\\[2ex]
    &+\frac{h_{\omega,k}^2 h_{\sigma,k}}{4\pi^2} \bar{m}^{*2}_N \Big[ \mathcal{FB}^a_{(2,2)}(\bar{m}^{*2}_N,\bar{m}_\omega^2;T,\mu^*_p,p_0)\nonumber\\[2ex]
    &\hspace{2cm}+2\mathcal{FB}^a_{(3,1)}(\bar{m}^{*2}_N,\bar{m}_\omega^2;T,\mu^*_p,p_0) \Big]\nonumber\\[2ex]
    &+(p\rightarrow n)\,.
\end{align}
The third part of the flow in \Eq{eq:dthsigma} reads
\begin{align}
    &\mathrm{Flow}_{h_\sigma,3}\nonumber\\[2ex]
    =&-\frac{3}{8\pi^2}h_{\rho,k}^2 h_{\sigma,k}\Big[\mathcal{FB}^a_{(2,1)}(\bar{m}^{*2}_N,\bar{m}_\rho^2;T,\mu^*_p,p_0)\nonumber\\[2ex]
    &\hspace{2cm}+\mathcal{FB}^a_{(1,2)}(\bar{m}^{*2}_N,\bar{m}_\rho^2;T,\mu^*_p,p_0) \Big]\nonumber\\[2ex]
    &+\frac{3}{4\pi^2}h_{\rho,k}^2 h_{\sigma,k} \bar{m}^{*2}_N \Big[ \mathcal{FB}^a_{(2,2)}(\bar{m}^{*2}_N,\bar{m}_\rho^2;T,\mu^*_p,p_0)\nonumber\\[2ex]
    &\hspace{2.5cm}+2\mathcal{FB}^a_{(3,1)}(\bar{m}^{*2}_N,\bar{m}_\rho^2;T,\mu^*_p,p_0) \Big]\nonumber\\[2ex]
    &+(p\rightarrow n)\,.
\end{align}
The fourth part of the flow of $h_{\sigma,k}$ in \Eq{eq:dthsigma} reads
\begin{align}
    &\mathrm{Flow}_{h_\sigma,4}\nonumber\\[2ex]
    =&\frac{h_{\sigma,k}^2}{12\pi^2}\frac{(\lambda_3+\lambda_4\sigma)}{k}\bar{m}^{*2}_N \Big[ \mathcal{FB}^a_{(2,2)}(\bar{m}^{*2}_N,\bar{m}_\sigma^2;T,\mu^*_p,p_0) \nonumber\\[2ex]
    &\hspace{2.5cm}+2\mathcal{FB}^a_{(1,3)}(\bar{m}^{*2}_N,\bar{m}_\sigma^2;T,\mu^*_p,p_0) \Big] \nonumber\\[2ex]
    &+(p\rightarrow n)\,.
\end{align}

In the same way, the flow of nucleon-$\omega$ coupling receives contributions from the three diagrams as shown in \Fig{fig:nucleonomega-equ}, to wit, 
\begin{align}
    \partial_t h_{\omega,k}=\mathrm{Flow}_{h_\omega,1}+\mathrm{Flow}_{h_\omega,2}+\mathrm{Flow}_{h_\omega,3}\,.\label{eq:dthomega}
\end{align}
One finds for the first part
\begin{align}
    &\mathrm{Flow}_{h_\omega,1}\nonumber\\[2ex]
    =&-\frac{1}{72\pi^2}h_{\sigma,k}^2 h_{\omega,k} \Big[5\mathcal{FB}^a_{(2,1)}(\bar{m}^{*2}_N,\bar{m}_\sigma^2;T,\mu^*_p,p_0) \nonumber\\[2ex]
    &\hspace{2.5cm}-4\mathcal{FB}^a_{(3,1)}(\bar{m}^{*2}_N,\bar{m}_\sigma^2;T,\mu^*_p,p_0)\nonumber\\[2ex]
    &\hspace{2.5cm}+3\mathcal{FB}^a_{(1,2)}(\bar{m}^{*2}_N,\bar{m}_\sigma^2;T,\mu^*_p,p_0)\nonumber\\[2ex]
    &\hspace{2.5cm}-2\mathcal{FB}^a_{(2,2)}(\bar{m}^{*2}_N,\bar{m}_\sigma^2;T,\mu^*_p,p_0) \Big]\nonumber\\[2ex]
    &+(p\rightarrow n)\,.
\end{align}
The second part of the flow of $h_{\omega,k}$ in \Eq{eq:dthomega} reads
\begin{align}
    &\mathrm{Flow}_{h_\omega,2}\nonumber\\[2ex]
    =&-\frac{h_{\omega,k}^3}{72\pi^2} \Big[5\mathcal{FB}^a_{(2,1)}(\bar{m}^{*2}_N,\bar{m}_\omega^2;T,\mu^*_p,p_0) \nonumber\\[2ex]
    &\hspace{1.2cm}-4\mathcal{FB}^a_{(3,1)}(\bar{m}^{*2}_N,\bar{m}_\omega^2;T,\mu^*_p,p_0)\nonumber\\[2ex]
    &\hspace{1.2cm}+3\mathcal{FB}^a_{(1,2)}(\bar{m}^{*2}_N,\bar{m}_\omega^2;T,\mu^*_p,p_0)\nonumber\\[2ex]
    &\hspace{1.2cm}-2\mathcal{FB}^a_{(2,2)}(\bar{m}^{*2}_N,\bar{m}_\omega^2;T,\mu^*_p,p_0) \Big]\nonumber\\[2ex]
    &-\frac{h_{\omega,k}^3}{24\pi^2}  \int_0^1 \mathrm{d}x \,x \Big[ \mathcal{FB}^c_{(2,1)}(\bar{m}^{*2}_N,\bar{m}_\omega^2,x;T,\mu^*_p,p_0)\nonumber\\[2ex]
    &\hspace{1.2cm}+x^{1/2} \Big(3\mathcal{FB}^b_{(2,1)}(\bar{m}^{*2}_N,\bar{m}_\omega^2,x;T,\mu^*_p,p_0)\nonumber\\[2ex]
    &\hspace{2cm}-4\mathcal{FB}^b_{(3,1)}(\bar{m}^{*2}_N,\bar{m}_\omega^2,x;T,\mu^*_p,p_0)\nonumber\\[2ex]
    &\hspace{2cm}+\mathcal{FB}^b_{(1,2)}(\bar{m}^{*2}_N,\bar{m}_\omega^2,x;T,\mu^*_p,p_0) \nonumber\\[2ex]
    &\hspace{2cm}-2\mathcal{FB}^b_{(2,2)}(\bar{m}^{*2}_N,\bar{m}_\omega^2,x;T,\mu^*_p,p_0) \Big)\nonumber\\[2ex]
    &\hspace{1.2cm}-2\Big( 2\mathcal{FB}^c_{(3,1)}(\bar{m}^{*2}_N,\bar{m}_\omega^2,x;T,\mu^*_p,p_0)\nonumber\\[2ex]
    &\hspace{2cm}+\mathcal{FB}^c_{(2,2)}(\bar{m}^{*2}_N,\bar{m}_\omega^2,x;T,\mu^*_p,p_0) \Big) \Big]\nonumber\\[2ex]
    &+(p\rightarrow n)\,.
\end{align}
The third part of the flow of $h_{\omega,k}$ in \Eq{eq:dthomega} reads
\begin{align}
    &\mathrm{Flow}_{h_\omega,3}\nonumber\\[2ex]
    =&-\frac{h_{\rho,k}^2h_{\omega,k}}{24\pi^2}  \Big[5\mathcal{FB}^a_{(2,1)}(\bar{m}^{*2}_N,\bar{m}_\rho^2;T,\mu^*_p,p_0) \nonumber\\[2ex]
    &\hspace{1.8cm}-4\mathcal{FB}^a_{(3,1)}(\bar{m}^{*2}_N,\bar{m}_\rho^2;T,\mu^*_p,p_0)\nonumber\\[2ex]
    &\hspace{1.8cm}+3\mathcal{FB}^a_{(1,2)}(\bar{m}^{*2}_N,\bar{m}_\rho^2;T,\mu^*_p,p_0)\nonumber\\[2ex]
    &\hspace{1.8cm}-2\mathcal{FB}^a_{(2,2)}(\bar{m}^{*2}_N,\bar{m}_\rho^2;T,\mu^*_p,p_0) \Big]\nonumber\\[2ex]
    &-\frac{h_{\rho,k}^2 h_{\omega,k}}{8\pi^2}  \int_0^1 \mathrm{d}x \,x \Big[ \mathcal{FB}^c_{(2,1)}(\bar{m}^{*2}_N,\bar{m}_\rho^2,x;T,\mu^*_p,p_0) \nonumber\\[2ex]
    &\hspace{1.8cm}+x^{1/2} \Big(3\mathcal{FB}^b_{(2,1)}(\bar{m}^{*2}_N,\bar{m}_\rho^2,x;T,\mu^*_p,p_0) \nonumber\\[2ex]
    &\hspace{2.5cm}-4\mathcal{FB}^b_{(3,1)}(\bar{m}^{*2}_N,\bar{m}_\rho^2,x;T,\mu^*_p,p_0) \nonumber\\[2ex]
    &\hspace{2.5cm}+\mathcal{FB}^b_{(1,2)}(\bar{m}^{*2}_N,\bar{m}_\rho^2,x;T,\mu^*_p,p_0) \nonumber\\[2ex]
    &\hspace{2.5cm}-2\mathcal{FB}^b_{(2,2)}(\bar{m}^{*2}_N,\bar{m}_\rho^2,x;T,\mu^*_p,p_0) \Big) \nonumber\\[2ex]
    &\hspace{1.8cm}-2\Big( 2\mathcal{FB}^c_{(3,1)}(\bar{m}^{*2}_N,\bar{m}_\rho^2,x;T,\mu^*_p,p_0) \nonumber\\[2ex]
    &\hspace{2.5cm}+\mathcal{FB}^c_{(2,2)}(\bar{m}^{*2}_N,\bar{m}_\rho^2,x;T,\mu^*_p,p_0) \Big) \Big]\nonumber\\[2ex]
    &+(p\rightarrow n) \,.
\end{align}
The flow of $h_{\rho,k}$ can also be written as
\begin{align}
    \partial_t h_{\rho,k}=\mathrm{Flow}_{h_\rho,1}+\mathrm{Flow}_{h_\rho,2}+\mathrm{Flow}_{h_\rho,3}\,,
\end{align}
which is in form related to the flow of $h_{\omega,k}$ through the relations as follows
\begin{align}
    \mathrm{Flow}_{h_\rho,1}=&\frac{1}{12} \frac{h_{\rho,k}}{h_{\omega,k}}\mathrm{Flow}_{h_\omega,1}\,,\\[2ex]
    \mathrm{Flow}_{h_\rho,2}=&\frac{1}{12} \frac{h_{\rho,k}}{h_{\omega,k}}\mathrm{Flow}_{h_\omega,2}\,,\\[2ex]
    \mathrm{Flow}_{h_\rho,3}=&-\frac{1}{36} \frac{h_{\rho,k}}{h_{\omega,k}}\mathrm{Flow}_{h_\omega,3}\,.
\end{align}

\section{Threshold functions}
\label{app:thres}

We define the dimensionless scalar propagators for the bosonic and fermionic fields
\begin{align}
    G_{b}(q,\bar{m}_{b}^{2})=&\frac{1}{\bar{q}_0^2+1+\bar{m}_{b}^{2}}\,,\\[2ex]
    G_{f}(q,\bar{m}_{f}^{2})=&\frac{1}{(\bar{q}_0+\mathrm{i}\bar{\mu})^2+1+\bar{m}_{f}^{2}}\,,\label{}
\end{align}
with $\bar{q}_0=q_0/k$, $\bar{\mu}=\mu/k$, $\bar{m}_{b}=m_{b}/k$ and $\bar{m}_{f}=m_{f}/k$. The Matsubara frequency reads $q_0=2n_q\pi T$ ($n_q\in \mathbb{Z}$) for bosons and $(2n_q+1)\pi T$ for fermions. To proceed, we define the threshold function as follows
\begin{align}
    \mathcal{F}_{(n)}(\bar{m}_{f}^{2};T,\mu)=\frac{T}{k}\sum_{n_q}\big(G_{f}(q,\bar{m}_{f}^{2})\big)^n\,.\label{eq:Fn}
\end{align}
A direct calculation by summing up the Matsubara frequencies leads us to
\begin{align}
    \mathcal{F}_{(1)}(\bar{m}_{f}^{2};T,\mu)=&\frac{1}{2\sqrt{1+\bar{m}_{f}^{2}}}\bigg(1-n_{F}(\bar{m}_{f}^{2};T,\mu)\nonumber\\[2ex]
    &-n_{F}(\bar{m}_{f}^{2};T,-\mu)\bigg)\,.\label{eq:F1}
\end{align}

Furthermore, we also need threshold functions mixing the fermionic and bosonic propagators, and there are three different types in the flows of Yukawa couplings, which read
\begin{align}
  &\mathcal{FB}^a_{(n_f,n_b)}(\bar m_f^2,\bar m_b^2;T,\mu,p_0)\nonumber\\[2ex]
  =&\frac{T}{k}\sum_{n_q}\Big(G_{f}(q,\bar m_f^2)\Big)^{n_f}\Big(G_{b}(q-p,\bar m_b^2)\Big)^{n_b}\,,\\[2ex]
  &\mathcal{FB}^b_{(n_f,n_b)}(\bar m_f^2,\bar m_b^2,x=\bar{\bm q}^2;T,\mu,p_0)\nonumber\\[2ex]
  =&\frac{T}{k}\sum_{n_q}\Big(G_{f}(q,\bar m_f^2)\Big)^{n_f}\Big(G_{b}(q-p,\bar m_b^2)\Big)^{n_b}\nonumber\\[2ex]
  &\hspace{0.2cm}\times\frac{1}{(\bar{q}_0-\bar{p}_0)^2+\bar{\bm q}^2}\,,\\[2ex]
  &\mathcal{FB}^c_{(n_f,n_b)}(\bar m_f^2,\bar m_b^2,x=\bar{\bm q}^2;T,\mu,p_0)\nonumber\\[2ex]
  =&\frac{T}{k}\sum_{n_q}\Big(G_{f}(q,\bar m_f^2)\Big)^{n_f}\Big(G_{b}(q-p,\bar m_b^2)\Big)^{n_b}\nonumber\\[2ex]
  &\hspace{0.2cm}\times\frac{1}{(\bar{q}_0-\bar{p}_0)^2+\bar{\bm q}^2}(\bar{q}_0+\mathrm{i}\bar{\mu})(\bar{q}_0-\bar{p}_0)\,,\label{}
\end{align}
with $x=\bar{\bm q}^2=\bm q^2/k^2\in [0,1]$. It is convenient to obtain higher-order threshold functions from the lower-order ones, e.g.,
\begin{align}
  &\mathcal{FB}_{(n_f+1,n_b)}(m_f^2,m_b^2)=-\frac{1}{n_f}\frac{\partial}{\partial m_f^2}\mathcal{FB}_{(n_f,n_b)}(m_f^2,m_b^2)\,,
\end{align}
and
\begin{align}
  \mathcal{FB}_{(n_f,n_b+1)}(m_f^2,m_b^2)=-\frac{1}{n_b}\frac{\partial}{\partial m_b^2}\mathcal{FB}_{(n_f,n_b)}(m_f^2,m_b^2)\,.
\end{align}
Therefore, it is sufficient to present the explicit expressions for the lowest-order threshold functions. The mixing threshold function of first type is given by
\begin{widetext}
\begin{align}
&\mathcal{FB}^a_{(1,1)}(\bar{m}_f^2,\bar{m}_b^2;T,\mu,p_0)\nonumber\\[2ex]
=&\frac{k^2}{2} \bigg\{ -n_B(\bar{m}_b^2;T)\frac{1}{\sqrt{1+\bar{m}_b^2}}\frac{1}{\Big(\mathrm{i}p_0-\mu+k(1+\bar{m}_b^2)^{1/2}\Big)^2-k^2(1+\bar{m}_f^2)} \nonumber\\[2ex]
&\hspace{0.5cm}-\big(n_B(\bar{m}_b^2;T)+1\big)\frac{1}{\sqrt{1+\bar{m}_b^2}}\frac{1}{\Big(\mathrm{i}p_0-\mu-k(1+\bar{m}_b^2)^{1/2}\Big)^2-k^2(1+\bar{m}_f^2)} \nonumber\\[2ex]
&\hspace{0.5cm}+n_F(\bar{m}_f^2;T,-\mu)\frac{1}{\sqrt{1+\bar{m}_f^2}}\frac{1}{\Big(\mathrm{i}p_0-\mu-k(1+\bar{m}_f^2)^{1/2}\Big)^2-k^2(1+\bar{m}_b^2)} \nonumber\\[2ex]
&\hspace{0.5cm}+\big(n_F(\bar{m}_f^2;T,\mu)-1\big)\frac{1}{\sqrt{1+\bar{m}_f^2}}\frac{1}{\Big(\mathrm{i}p_0-\mu+k(1+\bar{m}_f^2)^{1/2}\Big)^2-k^2(1+\bar{m}_b^2)}\bigg\}\,, 
\end{align}
\end{widetext}
where the bosonic distribution function reads
\begin{align}
    n_{B}(\bar{m}_b^2;T)=\frac{1}{\exp\bigg\{\frac{k}{T}\Big(1+\bar{m}_b^2\Big)^{1/2}\bigg\}-1}\,.\label{eq:nB}
\end{align}
Moreover, we also need the bosonic distribution function with momentum $|\bm q|=kx$, i.e.,
\begin{align}
    n_{B}(\bar{m}_b^2,x;T)=\frac{1}{\exp\bigg\{\frac{k}{T}\Big(x+\bar{m}_b^2\Big)^{1/2}\bigg\}-1}\,,\label{eq:nB-x}
\end{align}
for the threshold functions in the following. The threshold function of second type reads
\begin{widetext}
\begin{align}
&\mathcal{FB}^b_{(1,1)}(\bar{m}_f^2,\bar{m}_b^2,x;T,\mu,p_0)\nonumber\\[2ex]
=&\frac{k^2}{2} \bigg\{ n_B(\bar{m}_b^2;T)\frac{1}{\sqrt{1+\bar{m}_b^2}} \frac{k^2}{\big(k^2(1+\bar{m}_b^2)-x k^2\big) \left(\Big(\mathrm{i}p_0-\mu+k(1+\bar{m}_b^2)^{1/2}\Big)^2-k^2(1+\bar{m}_f^2)\right)}\nonumber\\[2ex]
&\hspace{0.5cm}+(n_B(\bar{m}_b^2;T)+1) \frac{1}{\sqrt{1+\bar{m}_b^2}} \frac{k^2}{\big(k^2(1+\bar{m}_b^2)-x k^2\big) \left(\Big(\mathrm{i}p_0-\mu-k(1+\bar{m}_b^2)^{1/2}\Big)^2-k^2(1+\bar{m}_f^2)\right)}\nonumber\\[2ex]
&\hspace{0.5cm}-n_F(\bar{m}_f^2;T,-\mu)\frac{1}{\sqrt{1+\bar{m}_f^2}}\frac{k^2}{\left(k^2(1+\bar{m}_f^2)-x k^2 +(-\mathrm{i}p_0+\mu)\Big(2k(1+\bar{m}_f^2)^{1/2}-\mathrm{i}p_0+\mu\Big) \right)} \nonumber\\[2ex]
&\hspace{4cm}\times\frac{1}{\Big(\mathrm{i}p_0-\mu-k(1+\bar{m}_f^2)^{1/2}\Big)^2-k^2(1+\bar{m}_b^2)} \nonumber\\[2ex]
&\hspace{0.5cm}-(n_F(\bar{m}_f^2;T,\mu)-1)\frac{1}{\sqrt{1+\bar{m}_f^2}} \frac{k^2}{\left(k^2(1+\bar{m}_f^2)-x k^2 +(\mathrm{i}p_0-\mu)\Big(2k(1+\bar{m}_f^2)^{1/2}+\mathrm{i}p_0-\mu\Big) \right)}\nonumber\\[2ex]
&\hspace{4cm}\times\frac{1}{\Big(\mathrm{i}p_0-\mu+k(1+\bar{m}_f^2)^{1/2}\Big)^2-k^2(1+\bar{m}_b^2)}\nonumber\\[2ex]
&\hspace{0.5cm}+n_B(\bar{m}_b^2,x;T)\frac{k^3}{x^{1/2}k\Big(k^2(1+\bar{m}_b^2)-x k^2\Big)\left(k^2(1+\bar{m}_f^2)+\Big(p_0-\mathrm{i}(x^{1/2}k-\mu)\Big)^2\right)}\nonumber\\[2ex]
&\hspace{0.5cm}+(n_B(\bar{m}_b^2,x;T)+1)\frac{k^3}{x^{1/2}k\Big(k^2(1+\bar{m}_b^2)-x k^2\Big)\left(k^2(1+\bar{m}_f^2)+\Big(p_0+\mathrm{i}(x^{1/2}k+\mu)\Big)^2\right)}\bigg\}\,.
\end{align}
\end{widetext}
The threshold function of third type reads
\begin{widetext}
\begin{align}
    &\mathcal{FB}^c_{(1,1)}(\bar{m}_f^2,\bar{m}_b^2,x;T,\mu,p_0)\nonumber\\[2ex]
=&\frac{k^2}{2} \bigg\{ -n_B(\bar{m}_b^2;T)
\frac{k\left(k\sqrt{1+\bar{m}_b^2}+\mathrm{i}p_0-\mu\right)}{\left(k^2(1+\bar{m}_b^2)-x k^2\right)} \frac{1}{\left(\Big(\mathrm{i}p_0-\mu+k(1+\bar{m}_b^2)^{1/2}\Big)^2-k^2(1+\bar{m}_f^2)\right)}\nonumber\\[2ex]
&\hspace{0.5cm}-(n_B(\bar{m}_b^2;T)+1) \frac{k\left(k\sqrt{1+\bar{m}_b^2}-\mathrm{i}p_0+\mu\right)}{\left(k^2(1+\bar{m}_b^2)-x k^2\right)} \frac{1}{\left(\Big(\mathrm{i}p_0-\mu-k(1+\bar{m}_b^2)^{1/2}\Big)^2-k^2(1+\bar{m}_f^2)\right)}\nonumber\\[2ex]
&\hspace{0.5cm}+n_F(\bar{m}_f^2;T,-\mu) \frac{k\left(k\sqrt{1+\bar{m}_f^2}-\mathrm{i}p_0+\mu\right)}{\left(k^2(1+\bar{m}_f^2)-x k^2+(-\mathrm{i}p_0+\mu)\Big(2k(1+\bar{m}_f^2)^{1/2}-\mathrm{i}p_0+\mu\Big)\right)}\nonumber\\[2ex]
&\hspace{4cm}\times \frac{1}{\left(\Big(\mathrm{i}p_0-\mu-k(1+\bar{m}_f^2)^{1/2}\Big)^2-k^2(1+\bar{m}_b^2)\right)}\nonumber\\[2ex]
&\hspace{0.5cm}+(n_F(\bar{m}_f^2;T,-\mu)-1) \frac{k\left(k\sqrt{1+\bar{m}_f^2}+\mathrm{i}p_0-\mu\right)}{\left(k^2(1+\bar{m}_f^2)-x k^2+(\mathrm{i}p_0-\mu)\Big(2k(1+\bar{m}_f^2)^{1/2}+\mathrm{i}p_0-\mu\Big)\right)} \nonumber\\[2ex]
&\hspace{4cm}\times \frac{1}{\left(\Big(\mathrm{i}p_0-\mu+k(1+\bar{m}_f^2)^{1/2}\Big)^2-k^2(1+\bar{m}_b^2)\right)}\nonumber\\[2ex]
&\hspace{0.5cm}+n_B(\bar{m}_b^2,x;T) \frac{k(-\mathrm{i}p_0-x^{1/2}k+\mu)}{\left(k^2(1+\bar{m}_b^2)-x k^2\right) \left(k^2(1+\bar{m}_f^2)+(p_0-\mathrm{i}(x^{1/2}k-\mu))^2\right)}\nonumber\\[2ex]
&\hspace{0.5cm}+(n_B(\bar{m}_b^2,x;T)+1) \frac{-k(-\mathrm{i}p_0+x^{1/2}k+\mu)}{\left(k^2(1+\bar{m}_b^2)-x k^2\right) \left(k^2(1+\bar{m}_f^2)+(p_0+\mathrm{i}(x^{1/2}k+\mu))^2\right)}\bigg\}\,.
\end{align}
\end{widetext}

\vfill 

\bibliography{ref-lib}%

\begin{thebibliography}{72}%
\makeatletter
\providecommand \@ifxundefined [1]{%
 \@ifx{#1\undefined}
}%
\providecommand \@ifnum [1]{%
 \ifnum #1\expandafter \@firstoftwo
 \else \expandafter \@secondoftwo
 \fi
}%
\providecommand \@ifx [1]{%
 \ifx #1\expandafter \@firstoftwo
 \else \expandafter \@secondoftwo
 \fi
}%
\providecommand \natexlab [1]{#1}%
\providecommand \enquote  [1]{``#1''}%
\providecommand \bibnamefont  [1]{#1}%
\providecommand \bibfnamefont [1]{#1}%
\providecommand \citenamefont [1]{#1}%
\providecommand \href@noop [0]{\@secondoftwo}%
\providecommand \href [0]{\begingroup \@sanitize@url \@href}%
\providecommand \@href[1]{\@@startlink{#1}\@@href}%
\providecommand \@@href[1]{\endgroup#1\@@endlink}%
\providecommand \@sanitize@url [0]{\catcode `\\12\catcode `\$12\catcode
  `\&12\catcode `\#12\catcode `\^12\catcode `\_12\catcode `\%12\relax}%
\providecommand \@@startlink[1]{}%
\providecommand \@@endlink[0]{}%
\providecommand \url  [0]{\begingroup\@sanitize@url \@url }%
\providecommand \@url [1]{\endgroup\@href {#1}{\urlprefix }}%
\providecommand \urlprefix  [0]{URL }%
\providecommand \Eprint [0]{\href }%
\providecommand \doibase [0]{https://doi.org/}%
\providecommand \selectlanguage [0]{\@gobble}%
\providecommand \bibinfo  [0]{\@secondoftwo}%
\providecommand \bibfield  [0]{\@secondoftwo}%
\providecommand \translation [1]{[#1]}%
\providecommand \BibitemOpen [0]{}%
\providecommand \bibitemStop [0]{}%
\providecommand \bibitemNoStop [0]{.\EOS\space}%
\providecommand \EOS [0]{\spacefactor3000\relax}%
\providecommand \BibitemShut  [1]{\csname bibitem#1\endcsname}%
\let\auto@bib@innerbib\@empty
\bibitem [{\citenamefont {Stephanov}(2006)}]{Stephanov:2007fk}%
  \BibitemOpen
  \bibfield  {author} {\bibinfo {author} {\bibfnamefont {M.~A.}\ \bibnamefont
  {Stephanov}},\ }\bibfield  {title} {\bibinfo {title} {{QCD phase diagram: An
  Overview}},\ }\bibfield  {booktitle} {\emph {\bibinfo {booktitle}
  {{Proceedings, 24th International Symposium on Lattice Field Theory (Lattice
  2006): Tucson, USA, July 23-28, 2006}}},\ }\href@noop {} {\bibfield
  {journal} {\bibinfo  {journal} {PoS}\ }\textbf {\bibinfo {volume}
  {LAT2006}},\ \bibinfo {pages} {024} (\bibinfo {year} {2006})},\ \Eprint
  {https://arxiv.org/abs/hep-lat/0701002} {arXiv:hep-lat/0701002 [hep-lat]}
  \BibitemShut {NoStop}%
\bibitem [{\citenamefont {Aggarwal}\ \emph {et~al.}(2010)\citenamefont
  {Aggarwal} \emph {et~al.}}]{STAR:2010vob}%
  \BibitemOpen
  \bibfield  {author} {\bibinfo {author} {\bibfnamefont {M.~M.}\ \bibnamefont
  {Aggarwal}} \emph {et~al.} (\bibinfo {collaboration} {STAR}),\ }\bibfield
  {title} {\bibinfo {title} {{An Experimental Exploration of the QCD Phase
  Diagram: The Search for the Critical Point and the Onset of
  De-confinement}},\ }\href@noop {} {\  (\bibinfo {year} {2010})},\ \Eprint
  {https://arxiv.org/abs/1007.2613} {arXiv:1007.2613 [nucl-ex]} \BibitemShut
  {NoStop}%
\bibitem [{\citenamefont {Luo}\ and\ \citenamefont {Xu}(2017)}]{Luo:2017faz}%
  \BibitemOpen
  \bibfield  {author} {\bibinfo {author} {\bibfnamefont {X.}~\bibnamefont
  {Luo}}\ and\ \bibinfo {author} {\bibfnamefont {N.}~\bibnamefont {Xu}},\
  }\bibfield  {title} {\bibinfo {title} {{Search for the QCD Critical Point
  with Fluctuations of Conserved Quantities in Relativistic Heavy-Ion
  Collisions at RHIC : An Overview}},\ }\href
  {https://doi.org/10.1007/s41365-017-0257-0} {\bibfield  {journal} {\bibinfo
  {journal} {Nucl. Sci. Tech.}\ }\textbf {\bibinfo {volume} {28}},\ \bibinfo
  {pages} {112} (\bibinfo {year} {2017})},\ \Eprint
  {https://arxiv.org/abs/1701.02105} {arXiv:1701.02105 [nucl-ex]} \BibitemShut
  {NoStop}%
\bibitem [{\citenamefont {Bzdak}\ \emph {et~al.}(2020)\citenamefont {Bzdak},
  \citenamefont {Esumi}, \citenamefont {Koch}, \citenamefont {Liao},
  \citenamefont {Stephanov},\ and\ \citenamefont {Xu}}]{Bzdak:2019pkr}%
  \BibitemOpen
  \bibfield  {author} {\bibinfo {author} {\bibfnamefont {A.}~\bibnamefont
  {Bzdak}}, \bibinfo {author} {\bibfnamefont {S.}~\bibnamefont {Esumi}},
  \bibinfo {author} {\bibfnamefont {V.}~\bibnamefont {Koch}}, \bibinfo {author}
  {\bibfnamefont {J.}~\bibnamefont {Liao}}, \bibinfo {author} {\bibfnamefont
  {M.}~\bibnamefont {Stephanov}},\ and\ \bibinfo {author} {\bibfnamefont
  {N.}~\bibnamefont {Xu}},\ }\bibfield  {title} {\bibinfo {title} {{Mapping the
  Phases of Quantum Chromodynamics with Beam Energy Scan}},\ }\href
  {https://doi.org/10.1016/j.physrep.2020.01.005} {\bibfield  {journal}
  {\bibinfo  {journal} {Phys. Rept.}\ }\textbf {\bibinfo {volume} {853}},\
  \bibinfo {pages} {1} (\bibinfo {year} {2020})},\ \Eprint
  {https://arxiv.org/abs/1906.00936} {arXiv:1906.00936 [nucl-th]} \BibitemShut
  {NoStop}%
\bibitem [{\citenamefont {Fu}(2022)}]{Fu:2022gou}%
  \BibitemOpen
  \bibfield  {author} {\bibinfo {author} {\bibfnamefont {W.-j.}\ \bibnamefont
  {Fu}},\ }\bibfield  {title} {\bibinfo {title} {{QCD at finite temperature and
  density within the fRG approach: an overview}},\ }\href
  {https://doi.org/10.1088/1572-9494/ac86be} {\bibfield  {journal} {\bibinfo
  {journal} {Commun. Theor. Phys.}\ }\textbf {\bibinfo {volume} {74}},\
  \bibinfo {pages} {097304} (\bibinfo {year} {2022})},\ \Eprint
  {https://arxiv.org/abs/2205.00468} {arXiv:2205.00468 [hep-ph]} \BibitemShut
  {NoStop}%
\bibitem [{\citenamefont {Chen}\ \emph {et~al.}(2024)\citenamefont {Chen} \emph
  {et~al.}}]{Chen:2024aom}%
  \BibitemOpen
  \bibfield  {author} {\bibinfo {author} {\bibfnamefont {J.}~\bibnamefont
  {Chen}} \emph {et~al.},\ }\bibfield  {title} {\bibinfo {title} {{Properties
  of the QCD matter: review of selected results from the relativistic heavy ion
  collider beam energy scan (RHIC BES) program}},\ }\href
  {https://doi.org/10.1007/s41365-024-01591-2} {\bibfield  {journal} {\bibinfo
  {journal} {Nucl. Sci. Tech.}\ }\textbf {\bibinfo {volume} {35}},\ \bibinfo
  {pages} {214} (\bibinfo {year} {2024})},\ \Eprint
  {https://arxiv.org/abs/2407.02935} {arXiv:2407.02935 [nucl-ex]} \BibitemShut
  {NoStop}%
\bibitem [{\citenamefont {Adam}\ \emph {et~al.}(2021)\citenamefont {Adam} \emph
  {et~al.}}]{STAR:2020tga}%
  \BibitemOpen
  \bibfield  {author} {\bibinfo {author} {\bibfnamefont {J.}~\bibnamefont
  {Adam}} \emph {et~al.} (\bibinfo {collaboration} {STAR}),\ }\bibfield
  {title} {\bibinfo {title} {{Nonmonotonic Energy Dependence of Net-Proton
  Number Fluctuations}},\ }\href
  {https://doi.org/10.1103/PhysRevLett.126.092301} {\bibfield  {journal}
  {\bibinfo  {journal} {Phys. Rev. Lett.}\ }\textbf {\bibinfo {volume} {126}},\
  \bibinfo {pages} {092301} (\bibinfo {year} {2021})},\ \Eprint
  {https://arxiv.org/abs/2001.02852} {arXiv:2001.02852 [nucl-ex]} \BibitemShut
  {NoStop}%
\bibitem [{\citenamefont {Abdallah}\ \emph {et~al.}(2022)\citenamefont
  {Abdallah} \emph {et~al.}}]{STAR:2021fge}%
  \BibitemOpen
  \bibfield  {author} {\bibinfo {author} {\bibfnamefont {M.~S.}\ \bibnamefont
  {Abdallah}} \emph {et~al.} (\bibinfo {collaboration} {STAR}),\ }\bibfield
  {title} {\bibinfo {title} {{Measurements of Proton High Order Cumulants in
  $\sqrt{s_{_{\mathrm{NN}}}}$ = 3 GeV Au+Au Collisions and Implications for the
  QCD Critical Point}},\ }\href
  {https://doi.org/10.1103/PhysRevLett.128.202303} {\bibfield  {journal}
  {\bibinfo  {journal} {Phys. Rev. Lett.}\ }\textbf {\bibinfo {volume} {128}},\
  \bibinfo {pages} {202303} (\bibinfo {year} {2022})},\ \Eprint
  {https://arxiv.org/abs/2112.00240} {arXiv:2112.00240 [nucl-ex]} \BibitemShut
  {NoStop}%
\bibitem [{\citenamefont {Aboona}\ \emph {et~al.}(2023)\citenamefont {Aboona}
  \emph {et~al.}}]{STAR:2022vlo}%
  \BibitemOpen
  \bibfield  {author} {\bibinfo {author} {\bibfnamefont {B.}~\bibnamefont
  {Aboona}} \emph {et~al.} (\bibinfo {collaboration} {STAR}),\ }\bibfield
  {title} {\bibinfo {title} {{Beam Energy Dependence of Fifth and Sixth-Order
  Net-proton Number Fluctuations in Au+Au Collisions at RHIC}},\ }\href
  {https://doi.org/10.1103/PhysRevLett.130.082301} {\bibfield  {journal}
  {\bibinfo  {journal} {Phys. Rev. Lett.}\ }\textbf {\bibinfo {volume} {130}},\
  \bibinfo {pages} {082301} (\bibinfo {year} {2023})},\ \Eprint
  {https://arxiv.org/abs/2207.09837} {arXiv:2207.09837 [nucl-ex]} \BibitemShut
  {NoStop}%
\bibitem [{\citenamefont {Abdallah}\ \emph {et~al.}(2023)\citenamefont
  {Abdallah} \emph {et~al.}}]{STAR:2022etb}%
  \BibitemOpen
  \bibfield  {author} {\bibinfo {author} {\bibfnamefont {M.}~\bibnamefont
  {Abdallah}} \emph {et~al.} (\bibinfo {collaboration} {STAR}),\ }\bibfield
  {title} {\bibinfo {title} {{Higher-order cumulants and correlation functions
  of proton multiplicity distributions in sNN=3~GeV~Au+Au collisions at the
  RHIC STAR experiment}},\ }\href {https://doi.org/10.1103/PhysRevC.107.024908}
  {\bibfield  {journal} {\bibinfo  {journal} {Phys. Rev. C}\ }\textbf {\bibinfo
  {volume} {107}},\ \bibinfo {pages} {024908} (\bibinfo {year} {2023})},\
  \Eprint {https://arxiv.org/abs/2209.11940} {arXiv:2209.11940 [nucl-ex]}
  \BibitemShut {NoStop}%
\bibitem [{STA(2025)}]{STAR:2025zdq}%
  \BibitemOpen
  \bibfield  {title} {\bibinfo {title} {{Precision Measurement of (Net-)proton
  Number Fluctuations in Au+Au Collisions at RHIC}},\ }\href@noop {} {\
  (\bibinfo {year} {2025})},\ \Eprint {https://arxiv.org/abs/2504.00817}
  {arXiv:2504.00817 [nucl-ex]} \BibitemShut {NoStop}%
\bibitem [{\citenamefont {Fu}\ \emph {et~al.}(2020)\citenamefont {Fu},
  \citenamefont {Pawlowski},\ and\ \citenamefont {Rennecke}}]{Fu:2019hdw}%
  \BibitemOpen
  \bibfield  {author} {\bibinfo {author} {\bibfnamefont {W.-j.}\ \bibnamefont
  {Fu}}, \bibinfo {author} {\bibfnamefont {J.~M.}\ \bibnamefont {Pawlowski}},\
  and\ \bibinfo {author} {\bibfnamefont {F.}~\bibnamefont {Rennecke}},\
  }\bibfield  {title} {\bibinfo {title} {{QCD phase structure at finite
  temperature and density}},\ }\href
  {https://doi.org/10.1103/PhysRevD.101.054032} {\bibfield  {journal} {\bibinfo
   {journal} {Phys. Rev. D}\ }\textbf {\bibinfo {volume} {101}},\ \bibinfo
  {pages} {054032} (\bibinfo {year} {2020})},\ \Eprint
  {https://arxiv.org/abs/1909.02991} {arXiv:1909.02991 [hep-ph]} \BibitemShut
  {NoStop}%
\bibitem [{\citenamefont {Gao}\ and\ \citenamefont
  {Pawlowski}(2021)}]{Gao:2020fbl}%
  \BibitemOpen
  \bibfield  {author} {\bibinfo {author} {\bibfnamefont {F.}~\bibnamefont
  {Gao}}\ and\ \bibinfo {author} {\bibfnamefont {J.~M.}\ \bibnamefont
  {Pawlowski}},\ }\bibfield  {title} {\bibinfo {title} {{Chiral phase structure
  and critical end point in QCD}},\ }\href
  {https://doi.org/10.1016/j.physletb.2021.136584} {\bibfield  {journal}
  {\bibinfo  {journal} {Phys. Lett. B}\ }\textbf {\bibinfo {volume} {820}},\
  \bibinfo {pages} {136584} (\bibinfo {year} {2021})},\ \Eprint
  {https://arxiv.org/abs/2010.13705} {arXiv:2010.13705 [hep-ph]} \BibitemShut
  {NoStop}%
\bibitem [{\citenamefont {Gunkel}\ and\ \citenamefont
  {Fischer}(2021)}]{Gunkel:2021oya}%
  \BibitemOpen
  \bibfield  {author} {\bibinfo {author} {\bibfnamefont {P.~J.}\ \bibnamefont
  {Gunkel}}\ and\ \bibinfo {author} {\bibfnamefont {C.~S.}\ \bibnamefont
  {Fischer}},\ }\bibfield  {title} {\bibinfo {title} {{Locating the critical
  endpoint of QCD: Mesonic backcoupling effects}},\ }\href
  {https://doi.org/10.1103/PhysRevD.104.054022} {\bibfield  {journal} {\bibinfo
   {journal} {Phys. Rev. D}\ }\textbf {\bibinfo {volume} {104}},\ \bibinfo
  {pages} {054022} (\bibinfo {year} {2021})},\ \Eprint
  {https://arxiv.org/abs/2106.08356} {arXiv:2106.08356 [hep-ph]} \BibitemShut
  {NoStop}%
\bibitem [{\citenamefont {Clarke}\ \emph {et~al.}(2024)\citenamefont {Clarke},
  \citenamefont {Dimopoulos}, \citenamefont {Di~Renzo}, \citenamefont
  {Goswami}, \citenamefont {Schmidt}, \citenamefont {Singh},\ and\
  \citenamefont {Zambello}}]{Clarke:2024ugt}%
  \BibitemOpen
  \bibfield  {author} {\bibinfo {author} {\bibfnamefont {D.~A.}\ \bibnamefont
  {Clarke}}, \bibinfo {author} {\bibfnamefont {P.}~\bibnamefont {Dimopoulos}},
  \bibinfo {author} {\bibfnamefont {F.}~\bibnamefont {Di~Renzo}}, \bibinfo
  {author} {\bibfnamefont {J.}~\bibnamefont {Goswami}}, \bibinfo {author}
  {\bibfnamefont {C.}~\bibnamefont {Schmidt}}, \bibinfo {author} {\bibfnamefont
  {S.}~\bibnamefont {Singh}},\ and\ \bibinfo {author} {\bibfnamefont
  {K.}~\bibnamefont {Zambello}},\ }\bibfield  {title} {\bibinfo {title}
  {{Searching for the QCD critical endpoint using multi-point Pad\'e
  approximations}},\ }\href@noop {} {\  (\bibinfo {year} {2024})},\ \Eprint
  {https://arxiv.org/abs/2405.10196} {arXiv:2405.10196 [hep-lat]} \BibitemShut
  {NoStop}%
\bibitem [{\citenamefont {Sorensen}\ and\ \citenamefont
  {Sorensen}(2024)}]{Sorensen:2024mry}%
  \BibitemOpen
  \bibfield  {author} {\bibinfo {author} {\bibfnamefont {A.}~\bibnamefont
  {Sorensen}}\ and\ \bibinfo {author} {\bibfnamefont {P.}~\bibnamefont
  {Sorensen}},\ }\bibfield  {title} {\bibinfo {title} {{Locating the critical
  point for the hadron to quark-gluon plasma phase transition from finite-size
  scaling of proton cumulants in heavy-ion collisions}},\ }\href@noop {} {\
  (\bibinfo {year} {2024})},\ \Eprint {https://arxiv.org/abs/2405.10278}
  {arXiv:2405.10278 [nucl-th]} \BibitemShut {NoStop}%
\bibitem [{\citenamefont {Bluhm}\ \emph {et~al.}(2025)\citenamefont {Bluhm},
  \citenamefont {Fujimoto}, \citenamefont {McLerran},\ and\ \citenamefont
  {Nahrgang}}]{Bluhm:2024uhj}%
  \BibitemOpen
  \bibfield  {author} {\bibinfo {author} {\bibfnamefont {M.}~\bibnamefont
  {Bluhm}}, \bibinfo {author} {\bibfnamefont {Y.}~\bibnamefont {Fujimoto}},
  \bibinfo {author} {\bibfnamefont {L.}~\bibnamefont {McLerran}},\ and\
  \bibinfo {author} {\bibfnamefont {M.}~\bibnamefont {Nahrgang}},\ }\bibfield
  {title} {\bibinfo {title} {{Quark saturation in the QCD phase diagram}},\
  }\href {https://doi.org/10.1103/PhysRevC.111.044914} {\bibfield  {journal}
  {\bibinfo  {journal} {Phys. Rev. C}\ }\textbf {\bibinfo {volume} {111}},\
  \bibinfo {pages} {044914} (\bibinfo {year} {2025})},\ \Eprint
  {https://arxiv.org/abs/2409.12088} {arXiv:2409.12088 [nucl-th]} \BibitemShut
  {NoStop}%
\bibitem [{\citenamefont {Abbott}\ \emph {et~al.}(2017)\citenamefont {Abbott}
  \emph {et~al.}}]{LIGOScientific:2017vwq}%
  \BibitemOpen
  \bibfield  {author} {\bibinfo {author} {\bibfnamefont {B.~P.}\ \bibnamefont
  {Abbott}} \emph {et~al.} (\bibinfo {collaboration} {LIGO Scientific,
  Virgo}),\ }\bibfield  {title} {\bibinfo {title} {{GW170817: Observation of
  Gravitational Waves from a Binary Neutron Star Inspiral}},\ }\href
  {https://doi.org/10.1103/PhysRevLett.119.161101} {\bibfield  {journal}
  {\bibinfo  {journal} {Phys. Rev. Lett.}\ }\textbf {\bibinfo {volume} {119}},\
  \bibinfo {pages} {161101} (\bibinfo {year} {2017})},\ \Eprint
  {https://arxiv.org/abs/1710.05832} {arXiv:1710.05832 [gr-qc]} \BibitemShut
  {NoStop}%
\bibitem [{\citenamefont {Fujimoto}\ \emph {et~al.}(2020)\citenamefont
  {Fujimoto}, \citenamefont {Fukushima},\ and\ \citenamefont
  {Murase}}]{Fujimoto:2019hxv}%
  \BibitemOpen
  \bibfield  {author} {\bibinfo {author} {\bibfnamefont {Y.}~\bibnamefont
  {Fujimoto}}, \bibinfo {author} {\bibfnamefont {K.}~\bibnamefont
  {Fukushima}},\ and\ \bibinfo {author} {\bibfnamefont {K.}~\bibnamefont
  {Murase}},\ }\bibfield  {title} {\bibinfo {title} {{Mapping neutron star data
  to the equation of state using the deep neural network}},\ }\href
  {https://doi.org/10.1103/PhysRevD.101.054016} {\bibfield  {journal} {\bibinfo
   {journal} {Phys. Rev. D}\ }\textbf {\bibinfo {volume} {101}},\ \bibinfo
  {pages} {054016} (\bibinfo {year} {2020})},\ \Eprint
  {https://arxiv.org/abs/1903.03400} {arXiv:1903.03400 [nucl-th]} \BibitemShut
  {NoStop}%
\bibitem [{\citenamefont {Koliogiannis}\ \emph {et~al.}(2021)\citenamefont
  {Koliogiannis}, \citenamefont {Kanakis-Pegios},\ and\ \citenamefont
  {Moustakidis}}]{Koliogiannis:2021ijs}%
  \BibitemOpen
  \bibfield  {author} {\bibinfo {author} {\bibfnamefont {P.~S.}\ \bibnamefont
  {Koliogiannis}}, \bibinfo {author} {\bibfnamefont {A.}~\bibnamefont
  {Kanakis-Pegios}},\ and\ \bibinfo {author} {\bibfnamefont {C.~C.}\
  \bibnamefont {Moustakidis}},\ }\bibfield  {title} {\bibinfo {title} {{Neutron
  Stars and Gravitational Waves: The Key Role of Nuclear Equation of State}},\
  }\href {https://doi.org/10.3390/foundations1020017} {\bibfield  {journal}
  {\bibinfo  {journal} {Foundations}\ }\textbf {\bibinfo {volume} {1}},\
  \bibinfo {pages} {217} (\bibinfo {year} {2021})},\ \Eprint
  {https://arxiv.org/abs/2110.13557} {arXiv:2110.13557 [nucl-th]} \BibitemShut
  {NoStop}%
\bibitem [{\citenamefont {Fujimoto}\ \emph {et~al.}(2022)\citenamefont
  {Fujimoto}, \citenamefont {Fukushima}, \citenamefont {McLerran},\ and\
  \citenamefont {Praszalowicz}}]{Fujimoto:2022ohj}%
  \BibitemOpen
  \bibfield  {author} {\bibinfo {author} {\bibfnamefont {Y.}~\bibnamefont
  {Fujimoto}}, \bibinfo {author} {\bibfnamefont {K.}~\bibnamefont {Fukushima}},
  \bibinfo {author} {\bibfnamefont {L.~D.}\ \bibnamefont {McLerran}},\ and\
  \bibinfo {author} {\bibfnamefont {M.}~\bibnamefont {Praszalowicz}},\
  }\bibfield  {title} {\bibinfo {title} {{Trace Anomaly as Signature of
  Conformality in Neutron Stars}},\ }\href
  {https://doi.org/10.1103/PhysRevLett.129.252702} {\bibfield  {journal}
  {\bibinfo  {journal} {Phys. Rev. Lett.}\ }\textbf {\bibinfo {volume} {129}},\
  \bibinfo {pages} {252702} (\bibinfo {year} {2022})},\ \Eprint
  {https://arxiv.org/abs/2207.06753} {arXiv:2207.06753 [nucl-th]} \BibitemShut
  {NoStop}%
\bibitem [{\citenamefont {Fukushima}(2025)}]{Fukushima:2025ujk}%
  \BibitemOpen
  \bibfield  {author} {\bibinfo {author} {\bibfnamefont {K.}~\bibnamefont
  {Fukushima}},\ }\bibfield  {title} {\bibinfo {title} {{QCD phase diagram and
  astrophysical implications}},\ }\href
  {https://doi.org/10.1016/j.jspc.2025.100066} {\bibfield  {journal} {\bibinfo
  {journal} {J. Subatomic Part. Cosmol.}\ }\textbf {\bibinfo {volume} {3}},\
  \bibinfo {pages} {100066} (\bibinfo {year} {2025})},\ \Eprint
  {https://arxiv.org/abs/2501.01907} {arXiv:2501.01907 [hep-ph]} \BibitemShut
  {NoStop}%
\bibitem [{\citenamefont {Hu}\ \emph {et~al.}(2023)\citenamefont {Hu},
  \citenamefont {Wang}, \citenamefont {Shao}, \citenamefont {Zhou},
  \citenamefont {Ye}, \citenamefont {Zhao}, \citenamefont {Sun}, \citenamefont
  {Lu},\ and\ \citenamefont {Xu}}]{Hu:2023niz}%
  \BibitemOpen
  \bibfield  {author} {\bibinfo {author} {\bibfnamefont {D.}~\bibnamefont
  {Hu}}, \bibinfo {author} {\bibfnamefont {X.}~\bibnamefont {Wang}}, \bibinfo
  {author} {\bibfnamefont {M.}~\bibnamefont {Shao}}, \bibinfo {author}
  {\bibfnamefont {Y.}~\bibnamefont {Zhou}}, \bibinfo {author} {\bibfnamefont
  {S.}~\bibnamefont {Ye}}, \bibinfo {author} {\bibfnamefont {L.}~\bibnamefont
  {Zhao}}, \bibinfo {author} {\bibfnamefont {Y.}~\bibnamefont {Sun}}, \bibinfo
  {author} {\bibfnamefont {J.}~\bibnamefont {Lu}},\ and\ \bibinfo {author}
  {\bibfnamefont {H.}~\bibnamefont {Xu}},\ }\bibfield  {title} {\bibinfo
  {title} {{Design and performance testing of a T0 detector for the CSR
  External-target Experiment}},\ }\href
  {https://doi.org/10.1016/j.nima.2023.168773} {\bibfield  {journal} {\bibinfo
  {journal} {Nucl. Instrum. Meth. A}\ }\textbf {\bibinfo {volume} {1057}},\
  \bibinfo {pages} {168773} (\bibinfo {year} {2023})},\ \Eprint
  {https://arxiv.org/abs/2304.02944} {arXiv:2304.02944 [physics.ins-det]}
  \BibitemShut {NoStop}%
\bibitem [{\citenamefont {Friman}\ \emph {et~al.}(2011)\citenamefont {Friman},
  \citenamefont {Hohne}, \citenamefont {Knoll}, \citenamefont {Leupold},
  \citenamefont {Randrup}, \citenamefont {Rapp},\ and\ \citenamefont
  {Senger}}]{Friman:2011zz}%
  \BibitemOpen
  \bibfield  {author} {\bibinfo {author} {\bibfnamefont {B.}~\bibnamefont
  {Friman}}, \bibinfo {author} {\bibfnamefont {C.}~\bibnamefont {Hohne}},
  \bibinfo {author} {\bibfnamefont {J.}~\bibnamefont {Knoll}}, \bibinfo
  {author} {\bibfnamefont {S.}~\bibnamefont {Leupold}}, \bibinfo {author}
  {\bibfnamefont {J.}~\bibnamefont {Randrup}}, \bibinfo {author} {\bibfnamefont
  {R.}~\bibnamefont {Rapp}},\ and\ \bibinfo {author} {\bibfnamefont
  {P.}~\bibnamefont {Senger}},\ }\bibfield  {title} {\bibinfo {title} {{The CBM
  physics book: Compressed baryonic matter in laboratory experiments}},\ }\href
  {https://doi.org/10.1007/978-3-642-13293-3} {\bibfield  {journal} {\bibinfo
  {journal} {Lect. Notes Phys.}\ }\textbf {\bibinfo {volume} {814}},\ \bibinfo
  {pages} {pp.1} (\bibinfo {year} {2011})}\BibitemShut {NoStop}%
\bibitem [{\citenamefont {Agakishiev}\ \emph {et~al.}(2009)\citenamefont
  {Agakishiev} \emph {et~al.}}]{Agakishiev:2009am}%
  \BibitemOpen
  \bibfield  {author} {\bibinfo {author} {\bibfnamefont {G.}~\bibnamefont
  {Agakishiev}} \emph {et~al.} (\bibinfo {collaboration} {HADES}),\ }\bibfield
  {title} {\bibinfo {title} {{The High-Acceptance Dielectron Spectrometer
  HADES}},\ }\href {https://doi.org/10.1140/epja/i2009-10807-5} {\bibfield
  {journal} {\bibinfo  {journal} {Eur. Phys. J.}\ }\textbf {\bibinfo {volume}
  {A41}},\ \bibinfo {pages} {243} (\bibinfo {year} {2009})},\ \Eprint
  {https://arxiv.org/abs/0902.3478} {arXiv:0902.3478 [nucl-ex]} \BibitemShut
  {NoStop}%
\bibitem [{\citenamefont {Yang}\ \emph {et~al.}(2013)\citenamefont {Yang} \emph
  {et~al.}}]{Yang:2013yeb}%
  \BibitemOpen
  \bibfield  {author} {\bibinfo {author} {\bibfnamefont {J.~C.}\ \bibnamefont
  {Yang}} \emph {et~al.},\ }\bibfield  {title} {\bibinfo {title} {{High
  Intensity heavy ion Accelerator Facility (HIAF) in China}},\ }\bibfield
  {booktitle} {\emph {\bibinfo {booktitle} {{Proceedings, 16th International
  Conference on Electromagnetic Isotope Separators and Techniques Related to
  their Applications (EMIS 2012): Matsue, Japan, December 2-7, 2012}}},\ }\href
  {https://doi.org/10.1016/j.nimb.2013.08.046} {\bibfield  {journal} {\bibinfo
  {journal} {Nucl. Instrum. Meth.}\ }\textbf {\bibinfo {volume} {B317}},\
  \bibinfo {pages} {263} (\bibinfo {year} {2013})}\BibitemShut {NoStop}%
\bibitem [{\citenamefont {Sorin}\ \emph {et~al.}(2011)\citenamefont {Sorin},
  \citenamefont {Kekelidze}, \citenamefont {Kovalenko}, \citenamefont
  {Lednicky}, \citenamefont {Meshkov},\ and\ \citenamefont
  {Trubnikov}}]{Sorin:2011zz}%
  \BibitemOpen
  \bibfield  {author} {\bibinfo {author} {\bibfnamefont {A.}~\bibnamefont
  {Sorin}}, \bibinfo {author} {\bibfnamefont {V.}~\bibnamefont {Kekelidze}},
  \bibinfo {author} {\bibfnamefont {A.}~\bibnamefont {Kovalenko}}, \bibinfo
  {author} {\bibfnamefont {R.}~\bibnamefont {Lednicky}}, \bibinfo {author}
  {\bibfnamefont {I.}~\bibnamefont {Meshkov}},\ and\ \bibinfo {author}
  {\bibfnamefont {G.}~\bibnamefont {Trubnikov}},\ }\bibfield  {title} {\bibinfo
  {title} {{Heavy-ion program at NICA/MPD at JINR}},\ }\bibfield  {booktitle}
  {\emph {\bibinfo {booktitle} {{Proceedings, 4th International Conference on
  Hard and Electromagnetic Probes of High-Energy Nuclear Collisions (Hard
  Probes 2010): Eilat, Israel, October 10-15, 2010}}},\ }\href
  {https://doi.org/10.1016/j.nuclphysa.2011.02.118} {\bibfield  {journal}
  {\bibinfo  {journal} {Nucl. Phys.}\ }\textbf {\bibinfo {volume} {A855}},\
  \bibinfo {pages} {510} (\bibinfo {year} {2011})}\BibitemShut {NoStop}%
\bibitem [{\citenamefont {Sakaguchi}(2017)}]{Sakaguchi:2017ggo}%
  \BibitemOpen
  \bibfield  {author} {\bibinfo {author} {\bibfnamefont {T.}~\bibnamefont
  {Sakaguchi}} (\bibinfo {collaboration} {J-PARC-HI}),\ }\bibfield  {title}
  {\bibinfo {title} {{Study of high baryon density QCD matter at J-PARC-HI}},\
  }\bibfield  {booktitle} {\emph {\bibinfo {booktitle} {{Proceedings, 26th
  International Conference on Ultra-relativistic Nucleus-Nucleus Collisions
  (Quark Matter 2017): Chicago, Illinois, USA, February 5-11, 2017}}},\ }\href
  {https://doi.org/10.1016/j.nuclphysa.2017.05.081} {\bibfield  {journal}
  {\bibinfo  {journal} {Nucl. Phys.}\ }\textbf {\bibinfo {volume} {A967}},\
  \bibinfo {pages} {896} (\bibinfo {year} {2017})}\BibitemShut {NoStop}%
\bibitem [{\citenamefont {Bellwied}\ \emph {et~al.}(2015)\citenamefont
  {Bellwied}, \citenamefont {Borsanyi}, \citenamefont {Fodor}, \citenamefont
  {Guenther}, \citenamefont {Katz}, \citenamefont {Ratti},\ and\ \citenamefont
  {Szabo}}]{Bellwied:2015rza}%
  \BibitemOpen
  \bibfield  {author} {\bibinfo {author} {\bibfnamefont {R.}~\bibnamefont
  {Bellwied}}, \bibinfo {author} {\bibfnamefont {S.}~\bibnamefont {Borsanyi}},
  \bibinfo {author} {\bibfnamefont {Z.}~\bibnamefont {Fodor}}, \bibinfo
  {author} {\bibfnamefont {J.}~\bibnamefont {Guenther}}, \bibinfo {author}
  {\bibfnamefont {S.~D.}\ \bibnamefont {Katz}}, \bibinfo {author}
  {\bibfnamefont {C.}~\bibnamefont {Ratti}},\ and\ \bibinfo {author}
  {\bibfnamefont {K.~K.}\ \bibnamefont {Szabo}},\ }\bibfield  {title} {\bibinfo
  {title} {{The QCD phase diagram from analytic continuation}},\ }\href
  {https://doi.org/10.1016/j.physletb.2015.11.011} {\bibfield  {journal}
  {\bibinfo  {journal} {Phys. Lett.}\ }\textbf {\bibinfo {volume} {B751}},\
  \bibinfo {pages} {559} (\bibinfo {year} {2015})},\ \Eprint
  {https://arxiv.org/abs/1507.07510} {arXiv:1507.07510 [hep-lat]} \BibitemShut
  {NoStop}%
\bibitem [{\citenamefont {Bazavov}\ \emph
  {et~al.}(2017{\natexlab{a}})\citenamefont {Bazavov} \emph
  {et~al.}}]{Bazavov:2017dus}%
  \BibitemOpen
  \bibfield  {author} {\bibinfo {author} {\bibfnamefont {A.}~\bibnamefont
  {Bazavov}} \emph {et~al.},\ }\bibfield  {title} {\bibinfo {title} {{The QCD
  Equation of State to $\mathcal{O}(\mu_B^6)$ from Lattice QCD}},\ }\href
  {https://doi.org/10.1103/PhysRevD.95.054504} {\bibfield  {journal} {\bibinfo
  {journal} {Phys. Rev.}\ }\textbf {\bibinfo {volume} {D95}},\ \bibinfo {pages}
  {054504} (\bibinfo {year} {2017}{\natexlab{a}})},\ \Eprint
  {https://arxiv.org/abs/1701.04325} {arXiv:1701.04325 [hep-lat]} \BibitemShut
  {NoStop}%
\bibitem [{\citenamefont {Bazavov}\ \emph
  {et~al.}(2017{\natexlab{b}})\citenamefont {Bazavov} \emph
  {et~al.}}]{Bazavov:2017tot}%
  \BibitemOpen
  \bibfield  {author} {\bibinfo {author} {\bibfnamefont {A.}~\bibnamefont
  {Bazavov}} \emph {et~al.} (\bibinfo {collaboration} {HotQCD}),\ }\bibfield
  {title} {\bibinfo {title} {{Skewness and kurtosis of net baryon-number
  distributions at small values of the baryon chemical potential}},\ }\href
  {https://doi.org/10.1103/PhysRevD.96.074510} {\bibfield  {journal} {\bibinfo
  {journal} {Phys. Rev.}\ }\textbf {\bibinfo {volume} {D96}},\ \bibinfo {pages}
  {074510} (\bibinfo {year} {2017}{\natexlab{b}})},\ \Eprint
  {https://arxiv.org/abs/1708.04897} {arXiv:1708.04897 [hep-lat]} \BibitemShut
  {NoStop}%
\bibitem [{\citenamefont {Borsanyi}\ \emph {et~al.}(2018)\citenamefont
  {Borsanyi}, \citenamefont {Fodor}, \citenamefont {Guenther}, \citenamefont
  {Katz}, \citenamefont {Szabo}, \citenamefont {Pasztor}, \citenamefont
  {Portillo},\ and\ \citenamefont {Ratti}}]{Borsanyi:2018grb}%
  \BibitemOpen
  \bibfield  {author} {\bibinfo {author} {\bibfnamefont {S.}~\bibnamefont
  {Borsanyi}}, \bibinfo {author} {\bibfnamefont {Z.}~\bibnamefont {Fodor}},
  \bibinfo {author} {\bibfnamefont {J.~N.}\ \bibnamefont {Guenther}}, \bibinfo
  {author} {\bibfnamefont {S.~K.}\ \bibnamefont {Katz}}, \bibinfo {author}
  {\bibfnamefont {K.~K.}\ \bibnamefont {Szabo}}, \bibinfo {author}
  {\bibfnamefont {A.}~\bibnamefont {Pasztor}}, \bibinfo {author} {\bibfnamefont
  {I.}~\bibnamefont {Portillo}},\ and\ \bibinfo {author} {\bibfnamefont
  {C.}~\bibnamefont {Ratti}},\ }\bibfield  {title} {\bibinfo {title} {{Higher
  order fluctuations and correlations of conserved charges from lattice QCD}},\
  }\href {https://doi.org/10.1007/JHEP10(2018)205} {\bibfield  {journal}
  {\bibinfo  {journal} {JHEP}\ }\textbf {\bibinfo {volume} {10}},\ \bibinfo
  {pages} {205}},\ \Eprint {https://arxiv.org/abs/1805.04445} {arXiv:1805.04445
  [hep-lat]} \BibitemShut {NoStop}%
\bibitem [{\citenamefont {Bazavov}\ \emph {et~al.}(2019)\citenamefont {Bazavov}
  \emph {et~al.}}]{Bazavov:2018mes}%
  \BibitemOpen
  \bibfield  {author} {\bibinfo {author} {\bibfnamefont {A.}~\bibnamefont
  {Bazavov}} \emph {et~al.} (\bibinfo {collaboration} {HotQCD}),\ }\bibfield
  {title} {\bibinfo {title} {{Chiral crossover in QCD at zero and non-zero
  chemical potentials}},\ }\href
  {https://doi.org/10.1016/j.physletb.2019.05.013} {\bibfield  {journal}
  {\bibinfo  {journal} {Phys. Lett.}\ }\textbf {\bibinfo {volume} {B795}},\
  \bibinfo {pages} {15} (\bibinfo {year} {2019})},\ \Eprint
  {https://arxiv.org/abs/1812.08235} {arXiv:1812.08235 [hep-lat]} \BibitemShut
  {NoStop}%
\bibitem [{\citenamefont {Ding}\ \emph {et~al.}(2019)\citenamefont {Ding} \emph
  {et~al.}}]{Ding:2019prx}%
  \BibitemOpen
  \bibfield  {author} {\bibinfo {author} {\bibfnamefont {H.~T.}\ \bibnamefont
  {Ding}} \emph {et~al.},\ }\bibfield  {title} {\bibinfo {title} {{Chiral Phase
  Transition Temperature in ( 2+1 )-Flavor QCD}},\ }\href
  {https://doi.org/10.1103/PhysRevLett.123.062002} {\bibfield  {journal}
  {\bibinfo  {journal} {Phys. Rev. Lett.}\ }\textbf {\bibinfo {volume} {123}},\
  \bibinfo {pages} {062002} (\bibinfo {year} {2019})},\ \Eprint
  {https://arxiv.org/abs/1903.04801} {arXiv:1903.04801 [hep-lat]} \BibitemShut
  {NoStop}%
\bibitem [{\citenamefont {Ding}\ \emph {et~al.}(2021)\citenamefont {Ding},
  \citenamefont {Li}, \citenamefont {Mukherjee}, \citenamefont {Tomiya},
  \citenamefont {Wang},\ and\ \citenamefont {Zhang}}]{Ding:2020xlj}%
  \BibitemOpen
  \bibfield  {author} {\bibinfo {author} {\bibfnamefont {H.~T.}\ \bibnamefont
  {Ding}}, \bibinfo {author} {\bibfnamefont {S.~T.}\ \bibnamefont {Li}},
  \bibinfo {author} {\bibfnamefont {S.}~\bibnamefont {Mukherjee}}, \bibinfo
  {author} {\bibfnamefont {A.}~\bibnamefont {Tomiya}}, \bibinfo {author}
  {\bibfnamefont {X.~D.}\ \bibnamefont {Wang}},\ and\ \bibinfo {author}
  {\bibfnamefont {Y.}~\bibnamefont {Zhang}},\ }\bibfield  {title} {\bibinfo
  {title} {{Correlated Dirac Eigenvalues and Axial Anomaly in Chiral Symmetric
  QCD}},\ }\href {https://doi.org/10.1103/PhysRevLett.126.082001} {\bibfield
  {journal} {\bibinfo  {journal} {Phys. Rev. Lett.}\ }\textbf {\bibinfo
  {volume} {126}},\ \bibinfo {pages} {082001} (\bibinfo {year} {2021})},\
  \Eprint {https://arxiv.org/abs/2010.14836} {arXiv:2010.14836 [hep-lat]}
  \BibitemShut {NoStop}%
\bibitem [{\citenamefont {Bazavov}\ \emph {et~al.}(2020)\citenamefont {Bazavov}
  \emph {et~al.}}]{Bazavov:2020bjn}%
  \BibitemOpen
  \bibfield  {author} {\bibinfo {author} {\bibfnamefont {A.}~\bibnamefont
  {Bazavov}} \emph {et~al.},\ }\bibfield  {title} {\bibinfo {title} {{Skewness,
  kurtosis, and the fifth and sixth order cumulants of net baryon-number
  distributions from lattice QCD confront high-statistics STAR data}},\ }\href
  {https://doi.org/10.1103/PhysRevD.101.074502} {\bibfield  {journal} {\bibinfo
   {journal} {Phys. Rev. D}\ }\textbf {\bibinfo {volume} {101}},\ \bibinfo
  {pages} {074502} (\bibinfo {year} {2020})},\ \Eprint
  {https://arxiv.org/abs/2001.08530} {arXiv:2001.08530 [hep-lat]} \BibitemShut
  {NoStop}%
\bibitem [{\citenamefont {Borsanyi}\ \emph {et~al.}(2020)\citenamefont
  {Borsanyi}, \citenamefont {Fodor}, \citenamefont {Guenther}, \citenamefont
  {Kara}, \citenamefont {Katz}, \citenamefont {Parotto}, \citenamefont
  {Pasztor}, \citenamefont {Ratti},\ and\ \citenamefont
  {Szabo}}]{Borsanyi:2020fev}%
  \BibitemOpen
  \bibfield  {author} {\bibinfo {author} {\bibfnamefont {S.}~\bibnamefont
  {Borsanyi}}, \bibinfo {author} {\bibfnamefont {Z.}~\bibnamefont {Fodor}},
  \bibinfo {author} {\bibfnamefont {J.~N.}\ \bibnamefont {Guenther}}, \bibinfo
  {author} {\bibfnamefont {R.}~\bibnamefont {Kara}}, \bibinfo {author}
  {\bibfnamefont {S.~D.}\ \bibnamefont {Katz}}, \bibinfo {author}
  {\bibfnamefont {P.}~\bibnamefont {Parotto}}, \bibinfo {author} {\bibfnamefont
  {A.}~\bibnamefont {Pasztor}}, \bibinfo {author} {\bibfnamefont
  {C.}~\bibnamefont {Ratti}},\ and\ \bibinfo {author} {\bibfnamefont {K.~K.}\
  \bibnamefont {Szabo}},\ }\bibfield  {title} {\bibinfo {title} {{QCD Crossover
  at Finite Chemical Potential from Lattice Simulations}},\ }\href
  {https://doi.org/10.1103/PhysRevLett.125.052001} {\bibfield  {journal}
  {\bibinfo  {journal} {Phys. Rev. Lett.}\ }\textbf {\bibinfo {volume} {125}},\
  \bibinfo {pages} {052001} (\bibinfo {year} {2020})},\ \Eprint
  {https://arxiv.org/abs/2002.02821} {arXiv:2002.02821 [hep-lat]} \BibitemShut
  {NoStop}%
\bibitem [{\citenamefont {Braun}\ \emph {et~al.}(2020)\citenamefont {Braun},
  \citenamefont {Fu}, \citenamefont {Pawlowski}, \citenamefont {Rennecke},
  \citenamefont {Rosenbl\"uh},\ and\ \citenamefont {Yin}}]{Braun:2020ada}%
  \BibitemOpen
  \bibfield  {author} {\bibinfo {author} {\bibfnamefont {J.}~\bibnamefont
  {Braun}}, \bibinfo {author} {\bibfnamefont {W.-j.}\ \bibnamefont {Fu}},
  \bibinfo {author} {\bibfnamefont {J.~M.}\ \bibnamefont {Pawlowski}}, \bibinfo
  {author} {\bibfnamefont {F.}~\bibnamefont {Rennecke}}, \bibinfo {author}
  {\bibfnamefont {D.}~\bibnamefont {Rosenbl\"uh}},\ and\ \bibinfo {author}
  {\bibfnamefont {S.}~\bibnamefont {Yin}},\ }\bibfield  {title} {\bibinfo
  {title} {{Chiral susceptibility in ( 2+1 )-flavor QCD}},\ }\href
  {https://doi.org/10.1103/PhysRevD.102.056010} {\bibfield  {journal} {\bibinfo
   {journal} {Phys. Rev. D}\ }\textbf {\bibinfo {volume} {102}},\ \bibinfo
  {pages} {056010} (\bibinfo {year} {2020})},\ \Eprint
  {https://arxiv.org/abs/2003.13112} {arXiv:2003.13112 [hep-ph]} \BibitemShut
  {NoStop}%
\bibitem [{\citenamefont {Braun}\ \emph {et~al.}(2025)\citenamefont {Braun}
  \emph {et~al.}}]{Braun:2023qak}%
  \BibitemOpen
  \bibfield  {author} {\bibinfo {author} {\bibfnamefont {J.}~\bibnamefont
  {Braun}} \emph {et~al.},\ }\bibfield  {title} {\bibinfo {title} {{Soft modes
  in hot QCD matter}},\ }\href {https://doi.org/10.1103/PhysRevD.111.094010}
  {\bibfield  {journal} {\bibinfo  {journal} {Phys. Rev. D}\ }\textbf {\bibinfo
  {volume} {111}},\ \bibinfo {pages} {094010} (\bibinfo {year} {2025})},\
  \Eprint {https://arxiv.org/abs/2310.19853} {arXiv:2310.19853 [hep-ph]}
  \BibitemShut {NoStop}%
\bibitem [{\citenamefont {Fu}\ \emph {et~al.}(2025{\natexlab{a}})\citenamefont
  {Fu}, \citenamefont {Pawlowski}, \citenamefont {Pisarski}, \citenamefont
  {Rennecke}, \citenamefont {Wen},\ and\ \citenamefont {Yin}}]{Fu:2024rto}%
  \BibitemOpen
  \bibfield  {author} {\bibinfo {author} {\bibfnamefont {W.-j.}\ \bibnamefont
  {Fu}}, \bibinfo {author} {\bibfnamefont {J.~M.}\ \bibnamefont {Pawlowski}},
  \bibinfo {author} {\bibfnamefont {R.~D.}\ \bibnamefont {Pisarski}}, \bibinfo
  {author} {\bibfnamefont {F.}~\bibnamefont {Rennecke}}, \bibinfo {author}
  {\bibfnamefont {R.}~\bibnamefont {Wen}},\ and\ \bibinfo {author}
  {\bibfnamefont {S.}~\bibnamefont {Yin}},\ }\bibfield  {title} {\bibinfo
  {title} {{QCD moat regime and its real-time properties}},\ }\href
  {https://doi.org/10.1103/PhysRevD.111.094026} {\bibfield  {journal} {\bibinfo
   {journal} {Phys. Rev. D}\ }\textbf {\bibinfo {volume} {111}},\ \bibinfo
  {pages} {094026} (\bibinfo {year} {2025}{\natexlab{a}})},\ \Eprint
  {https://arxiv.org/abs/2412.15949} {arXiv:2412.15949 [hep-ph]} \BibitemShut
  {NoStop}%
\bibitem [{\citenamefont {Fu}\ \emph {et~al.}(2025{\natexlab{b}})\citenamefont
  {Fu}, \citenamefont {Luo}, \citenamefont {Pawlowski}, \citenamefont
  {Rennecke},\ and\ \citenamefont {Yin}}]{Fu:2023lcm}%
  \BibitemOpen
  \bibfield  {author} {\bibinfo {author} {\bibfnamefont {W.-j.}\ \bibnamefont
  {Fu}}, \bibinfo {author} {\bibfnamefont {X.}~\bibnamefont {Luo}}, \bibinfo
  {author} {\bibfnamefont {J.~M.}\ \bibnamefont {Pawlowski}}, \bibinfo {author}
  {\bibfnamefont {F.}~\bibnamefont {Rennecke}},\ and\ \bibinfo {author}
  {\bibfnamefont {S.}~\bibnamefont {Yin}},\ }\bibfield  {title} {\bibinfo
  {title} {{Ripples of the QCD critical point}},\ }\href
  {https://doi.org/10.1103/PhysRevD.111.L031502} {\bibfield  {journal}
  {\bibinfo  {journal} {Phys. Rev. D}\ }\textbf {\bibinfo {volume} {111}},\
  \bibinfo {pages} {L031502} (\bibinfo {year} {2025}{\natexlab{b}})},\ \Eprint
  {https://arxiv.org/abs/2308.15508} {arXiv:2308.15508 [hep-ph]} \BibitemShut
  {NoStop}%
\bibitem [{\citenamefont {Mitter}\ \emph {et~al.}(2015)\citenamefont {Mitter},
  \citenamefont {Pawlowski},\ and\ \citenamefont
  {Strodthoff}}]{Mitter:2014wpa}%
  \BibitemOpen
  \bibfield  {author} {\bibinfo {author} {\bibfnamefont {M.}~\bibnamefont
  {Mitter}}, \bibinfo {author} {\bibfnamefont {J.~M.}\ \bibnamefont
  {Pawlowski}},\ and\ \bibinfo {author} {\bibfnamefont {N.}~\bibnamefont
  {Strodthoff}},\ }\bibfield  {title} {\bibinfo {title} {{Chiral symmetry
  breaking in continuum QCD}},\ }\href
  {https://doi.org/10.1103/PhysRevD.91.054035} {\bibfield  {journal} {\bibinfo
  {journal} {Phys. Rev.}\ }\textbf {\bibinfo {volume} {D91}},\ \bibinfo {pages}
  {054035} (\bibinfo {year} {2015})},\ \Eprint
  {https://arxiv.org/abs/1411.7978} {arXiv:1411.7978 [hep-ph]} \BibitemShut
  {NoStop}%
\bibitem [{\citenamefont {Braun}\ \emph {et~al.}(2016)\citenamefont {Braun},
  \citenamefont {Fister}, \citenamefont {Pawlowski},\ and\ \citenamefont
  {Rennecke}}]{Braun:2014ata}%
  \BibitemOpen
  \bibfield  {author} {\bibinfo {author} {\bibfnamefont {J.}~\bibnamefont
  {Braun}}, \bibinfo {author} {\bibfnamefont {L.}~\bibnamefont {Fister}},
  \bibinfo {author} {\bibfnamefont {J.~M.}\ \bibnamefont {Pawlowski}},\ and\
  \bibinfo {author} {\bibfnamefont {F.}~\bibnamefont {Rennecke}},\ }\bibfield
  {title} {\bibinfo {title} {{From Quarks and Gluons to Hadrons: Chiral
  Symmetry Breaking in Dynamical QCD}},\ }\href
  {https://doi.org/10.1103/PhysRevD.94.034016} {\bibfield  {journal} {\bibinfo
  {journal} {Phys. Rev.}\ }\textbf {\bibinfo {volume} {D94}},\ \bibinfo {pages}
  {034016} (\bibinfo {year} {2016})},\ \Eprint
  {https://arxiv.org/abs/1412.1045} {arXiv:1412.1045 [hep-ph]} \BibitemShut
  {NoStop}%
\bibitem [{\citenamefont {Rennecke}(2015)}]{Rennecke:2015eba}%
  \BibitemOpen
  \bibfield  {author} {\bibinfo {author} {\bibfnamefont {F.}~\bibnamefont
  {Rennecke}},\ }\bibfield  {title} {\bibinfo {title} {{Vacuum structure of
  vector mesons in QCD}},\ }\href {https://doi.org/10.1103/PhysRevD.92.076012}
  {\bibfield  {journal} {\bibinfo  {journal} {Phys. Rev.}\ }\textbf {\bibinfo
  {volume} {D92}},\ \bibinfo {pages} {076012} (\bibinfo {year} {2015})},\
  \Eprint {https://arxiv.org/abs/1504.03585} {arXiv:1504.03585 [hep-ph]}
  \BibitemShut {NoStop}%
\bibitem [{\citenamefont {Cyrol}\ \emph {et~al.}(2016)\citenamefont {Cyrol},
  \citenamefont {Fister}, \citenamefont {Mitter}, \citenamefont {Pawlowski},\
  and\ \citenamefont {Strodthoff}}]{Cyrol:2016tym}%
  \BibitemOpen
  \bibfield  {author} {\bibinfo {author} {\bibfnamefont {A.~K.}\ \bibnamefont
  {Cyrol}}, \bibinfo {author} {\bibfnamefont {L.}~\bibnamefont {Fister}},
  \bibinfo {author} {\bibfnamefont {M.}~\bibnamefont {Mitter}}, \bibinfo
  {author} {\bibfnamefont {J.~M.}\ \bibnamefont {Pawlowski}},\ and\ \bibinfo
  {author} {\bibfnamefont {N.}~\bibnamefont {Strodthoff}},\ }\bibfield  {title}
  {\bibinfo {title} {{Landau gauge Yang-Mills correlation functions}},\ }\href
  {https://doi.org/10.1103/PhysRevD.94.054005} {\bibfield  {journal} {\bibinfo
  {journal} {Phys. Rev.}\ }\textbf {\bibinfo {volume} {D94}},\ \bibinfo {pages}
  {054005} (\bibinfo {year} {2016})},\ \Eprint
  {https://arxiv.org/abs/1605.01856} {arXiv:1605.01856 [hep-ph]} \BibitemShut
  {NoStop}%
\bibitem [{\citenamefont {Cyrol}\ \emph {et~al.}(2018)\citenamefont {Cyrol},
  \citenamefont {Mitter}, \citenamefont {Pawlowski},\ and\ \citenamefont
  {Strodthoff}}]{Cyrol:2017ewj}%
  \BibitemOpen
  \bibfield  {author} {\bibinfo {author} {\bibfnamefont {A.~K.}\ \bibnamefont
  {Cyrol}}, \bibinfo {author} {\bibfnamefont {M.}~\bibnamefont {Mitter}},
  \bibinfo {author} {\bibfnamefont {J.~M.}\ \bibnamefont {Pawlowski}},\ and\
  \bibinfo {author} {\bibfnamefont {N.}~\bibnamefont {Strodthoff}},\ }\bibfield
   {title} {\bibinfo {title} {{Nonperturbative quark, gluon, and meson
  correlators of unquenched QCD}},\ }\href
  {https://doi.org/10.1103/PhysRevD.97.054006} {\bibfield  {journal} {\bibinfo
  {journal} {Phys. Rev.}\ }\textbf {\bibinfo {volume} {D97}},\ \bibinfo {pages}
  {054006} (\bibinfo {year} {2018})},\ \Eprint
  {https://arxiv.org/abs/1706.06326} {arXiv:1706.06326 [hep-ph]} \BibitemShut
  {NoStop}%
\bibitem [{\citenamefont {Fu}\ \emph {et~al.}(2016)\citenamefont {Fu},
  \citenamefont {Pawlowski}, \citenamefont {Rennecke},\ and\ \citenamefont
  {Schaefer}}]{Fu:2016tey}%
  \BibitemOpen
  \bibfield  {author} {\bibinfo {author} {\bibfnamefont {W.-j.}\ \bibnamefont
  {Fu}}, \bibinfo {author} {\bibfnamefont {J.~M.}\ \bibnamefont {Pawlowski}},
  \bibinfo {author} {\bibfnamefont {F.}~\bibnamefont {Rennecke}},\ and\
  \bibinfo {author} {\bibfnamefont {B.-J.}\ \bibnamefont {Schaefer}},\
  }\bibfield  {title} {\bibinfo {title} {{Baryon number fluctuations at finite
  temperature and density}},\ }\href
  {https://doi.org/10.1103/PhysRevD.94.116020} {\bibfield  {journal} {\bibinfo
  {journal} {Phys. Rev. D}\ }\textbf {\bibinfo {volume} {94}},\ \bibinfo
  {pages} {116020} (\bibinfo {year} {2016})},\ \Eprint
  {https://arxiv.org/abs/1608.04302} {arXiv:1608.04302 [hep-ph]} \BibitemShut
  {NoStop}%
\bibitem [{\citenamefont {Fu}\ \emph {et~al.}(2021)\citenamefont {Fu},
  \citenamefont {Luo}, \citenamefont {Pawlowski}, \citenamefont {Rennecke},
  \citenamefont {Wen},\ and\ \citenamefont {Yin}}]{Fu:2021oaw}%
  \BibitemOpen
  \bibfield  {author} {\bibinfo {author} {\bibfnamefont {W.-j.}\ \bibnamefont
  {Fu}}, \bibinfo {author} {\bibfnamefont {X.}~\bibnamefont {Luo}}, \bibinfo
  {author} {\bibfnamefont {J.~M.}\ \bibnamefont {Pawlowski}}, \bibinfo {author}
  {\bibfnamefont {F.}~\bibnamefont {Rennecke}}, \bibinfo {author}
  {\bibfnamefont {R.}~\bibnamefont {Wen}},\ and\ \bibinfo {author}
  {\bibfnamefont {S.}~\bibnamefont {Yin}},\ }\bibfield  {title} {\bibinfo
  {title} {{Hyper-order baryon number fluctuations at finite temperature and
  density}},\ }\href {https://doi.org/10.1103/PhysRevD.104.094047} {\bibfield
  {journal} {\bibinfo  {journal} {Phys. Rev. D}\ }\textbf {\bibinfo {volume}
  {104}},\ \bibinfo {pages} {094047} (\bibinfo {year} {2021})},\ \Eprint
  {https://arxiv.org/abs/2101.06035} {arXiv:2101.06035 [hep-ph]} \BibitemShut
  {NoStop}%
\bibitem [{\citenamefont {Fu}\ \emph {et~al.}(2023)\citenamefont {Fu},
  \citenamefont {Huang}, \citenamefont {Pawlowski},\ and\ \citenamefont
  {Tan}}]{Fu:2022uow}%
  \BibitemOpen
  \bibfield  {author} {\bibinfo {author} {\bibfnamefont {W.-j.}\ \bibnamefont
  {Fu}}, \bibinfo {author} {\bibfnamefont {C.}~\bibnamefont {Huang}}, \bibinfo
  {author} {\bibfnamefont {J.~M.}\ \bibnamefont {Pawlowski}},\ and\ \bibinfo
  {author} {\bibfnamefont {Y.-y.}\ \bibnamefont {Tan}},\ }\bibfield  {title}
  {\bibinfo {title} {{Four-quark scatterings in QCD I}},\ }\href
  {https://doi.org/10.21468/SciPostPhys.14.4.069} {\bibfield  {journal}
  {\bibinfo  {journal} {SciPost Phys.}\ }\textbf {\bibinfo {volume} {14}},\
  \bibinfo {pages} {069} (\bibinfo {year} {2023})},\ \Eprint
  {https://arxiv.org/abs/2209.13120} {arXiv:2209.13120 [hep-ph]} \BibitemShut
  {NoStop}%
\bibitem [{\citenamefont {Fu}\ \emph {et~al.}(2024)\citenamefont {Fu},
  \citenamefont {Huang}, \citenamefont {Pawlowski},\ and\ \citenamefont
  {Tan}}]{Fu:2024ysj}%
  \BibitemOpen
  \bibfield  {author} {\bibinfo {author} {\bibfnamefont {W.-j.}\ \bibnamefont
  {Fu}}, \bibinfo {author} {\bibfnamefont {C.}~\bibnamefont {Huang}}, \bibinfo
  {author} {\bibfnamefont {J.~M.}\ \bibnamefont {Pawlowski}},\ and\ \bibinfo
  {author} {\bibfnamefont {Y.-y.}\ \bibnamefont {Tan}},\ }\bibfield  {title}
  {\bibinfo {title} {{Four-quark scatterings in QCD II}},\ }\href
  {https://doi.org/10.21468/SciPostPhys.17.5.148} {\bibfield  {journal}
  {\bibinfo  {journal} {SciPost Phys.}\ }\textbf {\bibinfo {volume} {17}},\
  \bibinfo {pages} {148} (\bibinfo {year} {2024})},\ \Eprint
  {https://arxiv.org/abs/2401.07638} {arXiv:2401.07638 [hep-ph]} \BibitemShut
  {NoStop}%
\bibitem [{\citenamefont {Tan}\ \emph {et~al.}(2025)\citenamefont {Tan},
  \citenamefont {Chen}, \citenamefont {Fu},\ and\ \citenamefont
  {Li}}]{Tan:2024fuq}%
  \BibitemOpen
  \bibfield  {author} {\bibinfo {author} {\bibfnamefont {Y.-y.}\ \bibnamefont
  {Tan}}, \bibinfo {author} {\bibfnamefont {Y.-r.}\ \bibnamefont {Chen}},
  \bibinfo {author} {\bibfnamefont {W.-j.}\ \bibnamefont {Fu}},\ and\ \bibinfo
  {author} {\bibfnamefont {W.-J.}\ \bibnamefont {Li}},\ }\bibfield  {title}
  {\bibinfo {title} {{Universality of pseudo-Goldstone damping near critical
  points}},\ }\href {https://doi.org/10.1038/s41467-025-58170-1} {\bibfield
  {journal} {\bibinfo  {journal} {Nature Commun.}\ }\textbf {\bibinfo {volume}
  {16}},\ \bibinfo {pages} {2916} (\bibinfo {year} {2025})},\ \Eprint
  {https://arxiv.org/abs/2403.03503} {arXiv:2403.03503 [hep-th]} \BibitemShut
  {NoStop}%
\bibitem [{\citenamefont {Ihssen}\ \emph {et~al.}(2024)\citenamefont {Ihssen},
  \citenamefont {Pawlowski}, \citenamefont {Sattler},\ and\ \citenamefont
  {Wink}}]{Ihssen:2024miv}%
  \BibitemOpen
  \bibfield  {author} {\bibinfo {author} {\bibfnamefont {F.}~\bibnamefont
  {Ihssen}}, \bibinfo {author} {\bibfnamefont {J.~M.}\ \bibnamefont
  {Pawlowski}}, \bibinfo {author} {\bibfnamefont {F.~R.}\ \bibnamefont
  {Sattler}},\ and\ \bibinfo {author} {\bibfnamefont {N.}~\bibnamefont
  {Wink}},\ }\bibfield  {title} {\bibinfo {title} {{Towards quantitative
  precision in functional QCD I}},\ }\href@noop {} {\  (\bibinfo {year}
  {2024})},\ \Eprint {https://arxiv.org/abs/2408.08413} {arXiv:2408.08413
  [hep-ph]} \BibitemShut {NoStop}%
\bibitem [{\citenamefont {Fu}\ \emph {et~al.}(2025{\natexlab{c}})\citenamefont
  {Fu}, \citenamefont {Huang}, \citenamefont {Pawlowski}, \citenamefont {Tan},\
  and\ \citenamefont {Zhou}}]{Fu:2025hcm}%
  \BibitemOpen
  \bibfield  {author} {\bibinfo {author} {\bibfnamefont {W.-j.}\ \bibnamefont
  {Fu}}, \bibinfo {author} {\bibfnamefont {C.}~\bibnamefont {Huang}}, \bibinfo
  {author} {\bibfnamefont {J.~M.}\ \bibnamefont {Pawlowski}}, \bibinfo {author}
  {\bibfnamefont {Y.-y.}\ \bibnamefont {Tan}},\ and\ \bibinfo {author}
  {\bibfnamefont {L.-j.}\ \bibnamefont {Zhou}},\ }\bibfield  {title} {\bibinfo
  {title} {{Four-quark scatterings in QCD III}},\ }\href@noop {} {\  (\bibinfo
  {year} {2025}{\natexlab{c}})},\ \Eprint {https://arxiv.org/abs/2502.14388}
  {arXiv:2502.14388 [hep-ph]} \BibitemShut {NoStop}%
\bibitem [{\citenamefont {Zhang}\ \emph {et~al.}(2025)\citenamefont {Zhang},
  \citenamefont {Huang},\ and\ \citenamefont {Fu}}]{Zhang:2025ofc}%
  \BibitemOpen
  \bibfield  {author} {\bibinfo {author} {\bibfnamefont {D.-y.}\ \bibnamefont
  {Zhang}}, \bibinfo {author} {\bibfnamefont {C.}~\bibnamefont {Huang}},\ and\
  \bibinfo {author} {\bibfnamefont {W.-j.}\ \bibnamefont {Fu}},\ }\bibfield
  {title} {\bibinfo {title} {{Quasi parton distributions of pions at large
  longitudinal momentum}},\ }\href@noop {} {\  (\bibinfo {year} {2025})},\
  \Eprint {https://arxiv.org/abs/2502.15384} {arXiv:2502.15384 [hep-ph]}
  \BibitemShut {NoStop}%
\bibitem [{\citenamefont {Chen}\ \emph
  {et~al.}(2025{\natexlab{a}})\citenamefont {Chen}, \citenamefont {Fu},
  \citenamefont {Yin},\ and\ \citenamefont {Zhang}}]{Chen:2025vwl}%
  \BibitemOpen
  \bibfield  {author} {\bibinfo {author} {\bibfnamefont {J.}~\bibnamefont
  {Chen}}, \bibinfo {author} {\bibfnamefont {W.-j.}\ \bibnamefont {Fu}},
  \bibinfo {author} {\bibfnamefont {S.}~\bibnamefont {Yin}},\ and\ \bibinfo
  {author} {\bibfnamefont {C.}~\bibnamefont {Zhang}},\ }\bibfield  {title}
  {\bibinfo {title} {{High-order fluctuations of temperature in hot QCD
  matter}},\ }\href@noop {} {\  (\bibinfo {year} {2025}{\natexlab{a}})},\
  \Eprint {https://arxiv.org/abs/2504.06886} {arXiv:2504.06886 [hep-ph]}
  \BibitemShut {NoStop}%
\bibitem [{\citenamefont {Gholami}\ \emph {et~al.}(2025)\citenamefont
  {Gholami}, \citenamefont {Kurth}, \citenamefont {Mire}, \citenamefont
  {Buballa},\ and\ \citenamefont {Schaefer}}]{Gholami:2025afm}%
  \BibitemOpen
  \bibfield  {author} {\bibinfo {author} {\bibfnamefont {H.}~\bibnamefont
  {Gholami}}, \bibinfo {author} {\bibfnamefont {L.}~\bibnamefont {Kurth}},
  \bibinfo {author} {\bibfnamefont {U.}~\bibnamefont {Mire}}, \bibinfo {author}
  {\bibfnamefont {M.}~\bibnamefont {Buballa}},\ and\ \bibinfo {author}
  {\bibfnamefont {B.-J.}\ \bibnamefont {Schaefer}},\ }\bibfield  {title}
  {\bibinfo {title} {{Renormalizing the Quark-Meson-Diquark Model}},\
  }\href@noop {} {\  (\bibinfo {year} {2025})},\ \Eprint
  {https://arxiv.org/abs/2505.22542} {arXiv:2505.22542 [hep-ph]} \BibitemShut
  {NoStop}%
\bibitem [{\citenamefont {Dupuis}\ \emph {et~al.}(2021)\citenamefont {Dupuis},
  \citenamefont {Canet}, \citenamefont {Eichhorn}, \citenamefont {Metzner},
  \citenamefont {Pawlowski}, \citenamefont {Tissier},\ and\ \citenamefont
  {Wschebor}}]{Dupuis:2020fhh}%
  \BibitemOpen
  \bibfield  {author} {\bibinfo {author} {\bibfnamefont {N.}~\bibnamefont
  {Dupuis}}, \bibinfo {author} {\bibfnamefont {L.}~\bibnamefont {Canet}},
  \bibinfo {author} {\bibfnamefont {A.}~\bibnamefont {Eichhorn}}, \bibinfo
  {author} {\bibfnamefont {W.}~\bibnamefont {Metzner}}, \bibinfo {author}
  {\bibfnamefont {J.~M.}\ \bibnamefont {Pawlowski}}, \bibinfo {author}
  {\bibfnamefont {M.}~\bibnamefont {Tissier}},\ and\ \bibinfo {author}
  {\bibfnamefont {N.}~\bibnamefont {Wschebor}},\ }\bibfield  {title} {\bibinfo
  {title} {{The nonperturbative functional renormalization group and its
  applications}},\ }\href {https://doi.org/10.1016/j.physrep.2021.01.001}
  {\bibfield  {journal} {\bibinfo  {journal} {Phys. Rept.}\ }\textbf {\bibinfo
  {volume} {910}},\ \bibinfo {pages} {1} (\bibinfo {year} {2021})},\ \Eprint
  {https://arxiv.org/abs/2006.04853} {arXiv:2006.04853 [cond-mat.stat-mech]}
  \BibitemShut {NoStop}%
\bibitem [{\citenamefont {Walecka}(1974)}]{Walecka:1974qa}%
  \BibitemOpen
  \bibfield  {author} {\bibinfo {author} {\bibfnamefont {J.~D.}\ \bibnamefont
  {Walecka}},\ }\bibfield  {title} {\bibinfo {title} {{A Theory of highly
  condensed matter}},\ }\href {https://doi.org/10.1016/0003-4916(74)90208-5}
  {\bibfield  {journal} {\bibinfo  {journal} {Annals Phys.}\ }\textbf {\bibinfo
  {volume} {83}},\ \bibinfo {pages} {491} (\bibinfo {year} {1974})}\BibitemShut
  {NoStop}%
\bibitem [{\citenamefont {Wetterich}(1993)}]{Wetterich:1992yh}%
  \BibitemOpen
  \bibfield  {author} {\bibinfo {author} {\bibfnamefont {C.}~\bibnamefont
  {Wetterich}},\ }\bibfield  {title} {\bibinfo {title} {{Exact evolution
  equation for the effective potential}},\ }\href
  {https://doi.org/10.1016/0370-2693(93)90726-X} {\bibfield  {journal}
  {\bibinfo  {journal} {Phys. Lett.}\ }\textbf {\bibinfo {volume} {B301}},\
  \bibinfo {pages} {90} (\bibinfo {year} {1993})}\BibitemShut {NoStop}%
\bibitem [{\citenamefont {Fu}\ \emph {et~al.}(2008{\natexlab{a}})\citenamefont
  {Fu}, \citenamefont {Wang},\ and\ \citenamefont {Liu}}]{Fu:2008zzg}%
  \BibitemOpen
  \bibfield  {author} {\bibinfo {author} {\bibfnamefont {W.-j.}\ \bibnamefont
  {Fu}}, \bibinfo {author} {\bibfnamefont {G.-h.}\ \bibnamefont {Wang}},\ and\
  \bibinfo {author} {\bibfnamefont {Y.-x.}\ \bibnamefont {Liu}},\ }\bibfield
  {title} {\bibinfo {title} {{Electron Capture and Its Reverse Process in Hot
  and Dense Astronuclear Matter}},\ }\href {https://doi.org/10.1086/528361}
  {\bibfield  {journal} {\bibinfo  {journal} {Astrophys. J.}\ }\textbf
  {\bibinfo {volume} {678}},\ \bibinfo {pages} {1517} (\bibinfo {year}
  {2008}{\natexlab{a}})}\BibitemShut {NoStop}%
\bibitem [{\citenamefont {Fu}\ \emph {et~al.}(2008{\natexlab{b}})\citenamefont
  {Fu}, \citenamefont {Wei},\ and\ \citenamefont {Liu}}]{Fu:2008bu}%
  \BibitemOpen
  \bibfield  {author} {\bibinfo {author} {\bibfnamefont {W.-j.}\ \bibnamefont
  {Fu}}, \bibinfo {author} {\bibfnamefont {H.-q.}\ \bibnamefont {Wei}},\ and\
  \bibinfo {author} {\bibfnamefont {Y.-x.}\ \bibnamefont {Liu}},\ }\bibfield
  {title} {\bibinfo {title} {{Distinguishing Newly Born Strange Stars from
  Neutron Stars with g-Mode Oscillations}},\ }\href
  {https://doi.org/10.1103/PhysRevLett.101.181102} {\bibfield  {journal}
  {\bibinfo  {journal} {Phys. Rev. Lett.}\ }\textbf {\bibinfo {volume} {101}},\
  \bibinfo {pages} {181102} (\bibinfo {year} {2008}{\natexlab{b}})},\ \Eprint
  {https://arxiv.org/abs/0810.1084} {arXiv:0810.1084 [nucl-th]} \BibitemShut
  {NoStop}%
\bibitem [{\citenamefont {Bai}\ \emph {et~al.}(2021)\citenamefont {Bai},
  \citenamefont {Fu},\ and\ \citenamefont {Liu}}]{Bai:2021wrh}%
  \BibitemOpen
  \bibfield  {author} {\bibinfo {author} {\bibfnamefont {Z.}~\bibnamefont
  {Bai}}, \bibinfo {author} {\bibfnamefont {W.-j.}\ \bibnamefont {Fu}},\ and\
  \bibinfo {author} {\bibfnamefont {Y.-x.}\ \bibnamefont {Liu}},\ }\bibfield
  {title} {\bibinfo {title} {{Identifying QCD Phase Transitions via the
  Gravitational Wave Frequency from a Supernova Explosion}},\ }\href
  {https://doi.org/10.3847/1538-4357/ac2a31} {\bibfield  {journal} {\bibinfo
  {journal} {Astrophys. J.}\ }\textbf {\bibinfo {volume} {922}},\ \bibinfo
  {pages} {266} (\bibinfo {year} {2021})},\ \Eprint
  {https://arxiv.org/abs/2109.12614} {arXiv:2109.12614 [nucl-th]} \BibitemShut
  {NoStop}%
\bibitem [{\citenamefont {Yin}\ \emph {et~al.}(2019)\citenamefont {Yin},
  \citenamefont {Wen},\ and\ \citenamefont {Fu}}]{Yin:2019ebz}%
  \BibitemOpen
  \bibfield  {author} {\bibinfo {author} {\bibfnamefont {S.}~\bibnamefont
  {Yin}}, \bibinfo {author} {\bibfnamefont {R.}~\bibnamefont {Wen}},\ and\
  \bibinfo {author} {\bibfnamefont {W.-j.}\ \bibnamefont {Fu}},\ }\bibfield
  {title} {\bibinfo {title} {{Mesonic dynamics and the QCD phase transition}},\
  }\href {https://doi.org/10.1103/PhysRevD.100.094029} {\bibfield  {journal}
  {\bibinfo  {journal} {Phys. Rev. D}\ }\textbf {\bibinfo {volume} {100}},\
  \bibinfo {pages} {094029} (\bibinfo {year} {2019})},\ \Eprint
  {https://arxiv.org/abs/1907.10262} {arXiv:1907.10262 [hep-ph]} \BibitemShut
  {NoStop}%
\bibitem [{\citenamefont {Cohen}(2003)}]{Cohen:2003kd}%
  \BibitemOpen
  \bibfield  {author} {\bibinfo {author} {\bibfnamefont {T.~D.~.}\ \bibnamefont
  {Cohen}},\ }\bibfield  {title} {\bibinfo {title} {{Functional integrals for
  QCD at nonzero chemical potential and zero density}},\ }\href
  {https://doi.org/10.1103/PhysRevLett.91.222001} {\bibfield  {journal}
  {\bibinfo  {journal} {Phys. Rev. Lett.}\ }\textbf {\bibinfo {volume} {91}},\
  \bibinfo {pages} {222001} (\bibinfo {year} {2003})},\ \Eprint
  {https://arxiv.org/abs/hep-ph/0307089} {arXiv:hep-ph/0307089} \BibitemShut
  {NoStop}%
\bibitem [{\citenamefont {Fu}\ and\ \citenamefont
  {Pawlowski}(2015)}]{Fu:2015naa}%
  \BibitemOpen
  \bibfield  {author} {\bibinfo {author} {\bibfnamefont {W.-j.}\ \bibnamefont
  {Fu}}\ and\ \bibinfo {author} {\bibfnamefont {J.~M.}\ \bibnamefont
  {Pawlowski}},\ }\bibfield  {title} {\bibinfo {title} {{Relevance of matter
  and glue dynamics for baryon number fluctuations}},\ }\href
  {https://doi.org/10.1103/PhysRevD.92.116006} {\bibfield  {journal} {\bibinfo
  {journal} {Phys. Rev.}\ }\textbf {\bibinfo {volume} {D92}},\ \bibinfo {pages}
  {116006} (\bibinfo {year} {2015})},\ \Eprint
  {https://arxiv.org/abs/1508.06504} {arXiv:1508.06504 [hep-ph]} \BibitemShut
  {NoStop}%
\bibitem [{\citenamefont {Navas}\ \emph {et~al.}(2024)\citenamefont {Navas}
  \emph {et~al.}}]{ParticleDataGroup:2024cfk}%
  \BibitemOpen
  \bibfield  {author} {\bibinfo {author} {\bibfnamefont {S.}~\bibnamefont
  {Navas}} \emph {et~al.} (\bibinfo {collaboration} {Particle Data Group}),\
  }\bibfield  {title} {\bibinfo {title} {{Review of particle physics}},\ }\href
  {https://doi.org/10.1103/PhysRevD.110.030001} {\bibfield  {journal} {\bibinfo
   {journal} {Phys. Rev. D}\ }\textbf {\bibinfo {volume} {110}},\ \bibinfo
  {pages} {030001} (\bibinfo {year} {2024})}\BibitemShut {NoStop}%
\bibitem [{\citenamefont {Lee}\ \emph {et~al.}(2015)\citenamefont {Lee},
  \citenamefont {Arrington},\ and\ \citenamefont {Hill}}]{Lee:2015jqa}%
  \BibitemOpen
  \bibfield  {author} {\bibinfo {author} {\bibfnamefont {G.}~\bibnamefont
  {Lee}}, \bibinfo {author} {\bibfnamefont {J.~R.}\ \bibnamefont {Arrington}},\
  and\ \bibinfo {author} {\bibfnamefont {R.~J.}\ \bibnamefont {Hill}},\
  }\bibfield  {title} {\bibinfo {title} {{Extraction of the proton radius from
  electron-proton scattering data}},\ }\href
  {https://doi.org/10.1103/PhysRevD.92.013013} {\bibfield  {journal} {\bibinfo
  {journal} {Phys. Rev. D}\ }\textbf {\bibinfo {volume} {92}},\ \bibinfo
  {pages} {013013} (\bibinfo {year} {2015})},\ \Eprint
  {https://arxiv.org/abs/1505.01489} {arXiv:1505.01489 [hep-ph]} \BibitemShut
  {NoStop}%
\bibitem [{\citenamefont {Bezginov}\ \emph {et~al.}(2019)\citenamefont
  {Bezginov}, \citenamefont {Valdez}, \citenamefont {Horbatsch}, \citenamefont
  {Marsman}, \citenamefont {Vutha},\ and\ \citenamefont
  {Hessels}}]{Bezginov:2019mdi}%
  \BibitemOpen
  \bibfield  {author} {\bibinfo {author} {\bibfnamefont {N.}~\bibnamefont
  {Bezginov}}, \bibinfo {author} {\bibfnamefont {T.}~\bibnamefont {Valdez}},
  \bibinfo {author} {\bibfnamefont {M.}~\bibnamefont {Horbatsch}}, \bibinfo
  {author} {\bibfnamefont {A.}~\bibnamefont {Marsman}}, \bibinfo {author}
  {\bibfnamefont {A.~C.}\ \bibnamefont {Vutha}},\ and\ \bibinfo {author}
  {\bibfnamefont {E.~A.}\ \bibnamefont {Hessels}},\ }\bibfield  {title}
  {\bibinfo {title} {{A measurement of the atomic hydrogen Lamb shift and the
  proton charge radius}},\ }\href {https://doi.org/10.1126/science.aau7807}
  {\bibfield  {journal} {\bibinfo  {journal} {Science}\ }\textbf {\bibinfo
  {volume} {365}},\ \bibinfo {pages} {1007} (\bibinfo {year}
  {2019})}\BibitemShut {NoStop}%
\bibitem [{\citenamefont {Chen}\ \emph
  {et~al.}(2025{\natexlab{b}})\citenamefont {Chen}, \citenamefont {Tan},
  \citenamefont {Fu}, \citenamefont {Luo},\ and\ \citenamefont
  {Chen}}]{Chen:2025a}%
  \BibitemOpen
  \bibfield  {author} {\bibinfo {author} {\bibfnamefont {Y.-r.}\ \bibnamefont
  {Chen}}, \bibinfo {author} {\bibfnamefont {Y.-y.}\ \bibnamefont {Tan}},
  \bibinfo {author} {\bibfnamefont {W.-j.}\ \bibnamefont {Fu}}, \bibinfo
  {author} {\bibfnamefont {X.}~\bibnamefont {Luo}},\ and\ \bibinfo {author}
  {\bibfnamefont {L.-W.}\ \bibnamefont {Chen}},\ }\href@noop {} {\bibfield
  {journal} {\bibinfo  {journal} {in preparation}\ } (\bibinfo {year}
  {2025}{\natexlab{b}})}\BibitemShut {NoStop}%
\bibitem [{\citenamefont {fQCD collaboration}()}]{fQCD}%
  \BibitemOpen
  \bibfield  {author} {\bibinfo {author} {\bibnamefont {fQCD collaboration}},\
  }\href {https://fqcd-collaboration.github.io} {\bibinfo  {journal}
  {https://fqcd-collaboration.github.io}\ }\BibitemShut {NoStop}%
\bibitem [{\citenamefont {Litim}(2000)}]{Litim:2000ci}%
  \BibitemOpen
\bibfield  {journal} {  }\bibfield  {author} {\bibinfo {author} {\bibfnamefont
  {D.~F.}\ \bibnamefont {Litim}},\ }\bibfield  {title} {\bibinfo {title}
  {{Optimization of the exact renormalization group}},\ }\href
  {https://doi.org/10.1016/S0370-2693(00)00748-6} {\bibfield  {journal}
  {\bibinfo  {journal} {Phys. Lett.}\ }\textbf {\bibinfo {volume} {B486}},\
  \bibinfo {pages} {92} (\bibinfo {year} {2000})},\ \Eprint
  {https://arxiv.org/abs/hep-th/0005245} {arXiv:hep-th/0005245 [hep-th]}
  \BibitemShut {NoStop}%
\bibitem [{\citenamefont {Litim}(2001)}]{Litim:2001up}%
  \BibitemOpen
  \bibfield  {author} {\bibinfo {author} {\bibfnamefont {D.~F.}\ \bibnamefont
  {Litim}},\ }\bibfield  {title} {\bibinfo {title} {{Optimized renormalization
  group flows}},\ }\href {https://doi.org/10.1103/PhysRevD.64.105007}
  {\bibfield  {journal} {\bibinfo  {journal} {Phys. Rev.}\ }\textbf {\bibinfo
  {volume} {D64}},\ \bibinfo {pages} {105007} (\bibinfo {year} {2001})},\
  \Eprint {https://arxiv.org/abs/hep-th/0103195} {arXiv:hep-th/0103195
  [hep-th]} \BibitemShut {NoStop}%
\end{thebibliography}%

\end{document}